\documentclass[aps,twocolumn,pra]{revtex4-1}
\usepackage{graphics}
\usepackage{dcolumn}% Align table columns on decimal point
\usepackage{bm}% bold math
\usepackage{amsmath}
\usepackage{hyperref}
\hypersetup{colorlinks=true, urlcolor=blue}
\usepackage{color}% color to the text 

\usepackage{newlfont}
\usepackage{amssymb}
\usepackage{amsfonts}
\usepackage{amsmath}
\usepackage{graphicx}
\usepackage{bm}

\def \be{\begin{equation}}
\def \ee{ \end{equation} }

\begin{document}

%opening
%\title{Quantum Correlation with Sandwiched Relative Entropies}
%\title{Quantum Correlation with Sandwiched Relative Entropies: Surpasses in scaling exponent compared to other order parameters in quantum phase transitions}
%\title{Quantum Correlation with Sandwiched Relative Entropies:\\ Better Order Parameter in Quantum Phase Transitions}
\title{Quantum Correlation with Sandwiched Relative Entropies:\\ Advantageous as Order Parameter in Quantum Phase Transitions}

\author{Avijit Misra, Anindya Biswas, Arun K. Pati, Aditi Sen(De), and Ujjwal Sen}

\affiliation{Harish-Chandra Research Institute, Chhatnag Road, Jhunsi, Allahabad 211 019, India}

\begin{abstract}

Quantum discord is a measure of quantum correlations beyond the 
entanglement-separability paradigm. It is conceptualized by using the von 
Neumann entropy as a measure of disorder. We introduce a class of quantum 
correlation measures as differences between total and classical 
correlations, in a shared quantum state, in terms of the sandwiched relative 
R\'{e}nyi and Tsallis entropies. 
We compare our results with those obtained by using the 
traditional relative entropies. We find that the measures satisfy all the plausible axioms for quantum correlations. 
We evaluate
the measures for  shared pure as well as paradigmatic classes of 
 mixed states. We show that the measures can faithfully detect the quantum critical point in 
the transverse quantum Ising model
%%%%%%%%%%%
and find that they can be 
used to remove an unquieting feature of nearest-neighbor quantum discord in this respect.
Furthermore, the 
measures provide better finite-size scaling  exponents of the quantum critical point than the ones for 
other known order parameters, including entanglement and information-theoretic measures of quantum correlations.
%, of the same quantum phase transition.

\end{abstract}

\maketitle

 \section{ Introduction}
 Characterization and quantification of quantum correlation~\cite{hor,kavan} play a central role in quantum information. 
 Entanglement, in particular, has been  successfully identified as a useful resource for 
 different quantum communication protocols~\cite{aus} and computational tasks~\cite{briegel}. 
 Moreover, it has also been employed to study cooperative
 quantum phenomena like quantum phase transitions in many-body systems~\cite{maciej,faziormp}. 
%  The measures in the entanglement-separability
%  paradigm include entanglement of formation~\cite{eof}, distillable entanglement~\cite{eof,distillable}, 
%  relative entropy of entanglement~\cite{VPRK}, logarithmic negativity~\cite{VidalWerner}, and many more~\cite{hor}.
 However, in the recent past, several quantum phenomena of shared systems have been discovered in which entanglement is either absent
 or does not play any significant role. 
 %These are the manifestations of quantum nonlocality without entanglement -- 
 Locally indistinguishable orthogonal product states~\cite{Chbennett3}~(c.f. \cite{Aperes}) is a 
 prominent example where entanglement
 does not play an important role.
 The role of entanglement is also unclear in the model of deterministic 
 quantum computation with one quantum bit~\cite{laflam,adatta,qc}. 
 %not required. 
 Such phenomena motivated the search for concepts and measures of quantum correlation independent of the 
 entanglement-separability paradigm.
% beyond the entanglement-separability paradigm which includes, to name a few, 
% entanglement of formation~\cite{eof}, distillable entanglement~\cite{eof,distillable}, 
% logarithmic negativity~\cite{VidalWerner} and relative entropy of entanglement~\cite{VPRK}. 
Introduction of quantum discord~\cite{hv,oz} is 
one of the most important advancements in this direction and has inspired a lot of
research activity~\cite{kavan}. It has thereby emerged that
%in which it was shown that 
quantum correlations, independent of entanglement, can also be a useful ingredient
in several quantum information processing tasks \cite{kavan}.
 Other measures in the same direction include quantum work deficit~\cite{workdeficit}, measurement-induced nonlocality~\cite{minonlcl}, 
and  quantum deficit~\cite{qdefi} (see also \cite{infprara}). These measures
 can be generally considered to be quantum correlation measures 
within an ``information-theoretic paradigm''.
%, quantifies
% like quantum work deficit, mesasures
%quantum correlations present in a bipartite system. Recently, the concept of quantum discord has been generalized 
%to the multipartite scenario using notions of dissonance~\cite{kavan1} and dissension~\cite{sir}.\\

In 
classical as well as quantum information theory, one of the most important pillars is the framework of 
entropy~\cite{wherl}, which 
%play an important role to 
quantifies the ignorance or lack of information in the relevant physical system.
%uncertainty. 
Moreover, it helps to understand information theory from a thermodynamic perspective. 
Almost all the quantum correlation measures incorporate entropic 
functions in various forms.
And, most  of the quantum correlation measures
%, mentioned earlier, 
are defined by using the von Neumann entropy~\cite{von}.
The operational significance of von Neumann entropy 
has been widely recognized in numerous scenarios in quantum information theory. Nonetheless,  there are classes 
of generalized entropies like the R\'{e}nyi~\cite{renyi} and Tsallis~\cite{tsallis} entropies, which are also operationally significant in 
important physical scenarios. 
Both the R\'{e}nyi and Tsallis entropies reduce to the von Neumann entropy when the
entropic parameter $\alpha$ $\rightarrow 1$.  
For $\alpha\in(0, 1),$  the relative R\'{e}nyi entropy appears in the quantum Chernoff 
bound which determines the minimal probability of error in discriminating two different 
quantum states in the setting of asymptotically many copies~\cite{chernf}. In Ref.~\cite{mosonyi}, it was shown that 
the relative R\'{e}nyi entropy is relevant in 
%the generalized cut-off rates in
binary quantum state discrimination, for the same range of $\alpha$.
The concept of R\'{e}nyi entropy has also been found to be useful in the context of holographic theory~\cite{holography}.
It has also been found useful in dealing with several condensed matter systems~\cite{conmat}. 
The significance of the Tsallis entropy in
quantum information theory has been established in the context of quantifying entanglement~\cite{abeakrqe},
local realism~\cite{abeakrlocal}, and entropic uncertainty relations~\cite{gunheml} (see also \cite{tsallisref}).
Both the R\'{e}nyi and Tsallis entropies have important applications in classical as well as quantum statistical mechanics and thermodynamics~\cite{statmech}.

While there are important interpretational and operational breakthroughs that have been obtained by using the concept of quantum discord,
there are also several intriguing unanswered questions and thriving controversies~\cite{kavan,dakic}. It is therefore interesting and
important to look back upon the conceptual foundations of quantum discord and inquire whether certain changes, subtle or substantial, in those
concepts lead us to a better understanding of 
%quantum correlations.
the controversies and the unanswered questions. 
Towards this aim,
%a hope in reaching this end, 
we introduce measures of the total, classical,
and quantum correlations of a bipartite quantum state in terms of the entire class of relative R\'{e}nyi and Tsallis entropy distances. 
%In particular,
%we use relative R\'{e}nyi and Tsallis entropies to define classical and quantum correlations. 
We show that the measures satisfy all the required properties 
of bipartite correlations. 
We then evaluate the quantum correlation measure for several paradigmatic classes of states. As an 
application, we find that the quantum correlation measures, via relative R\'{e}nyi and Tsallis entropies, can indicate quantum phase transitions
and give better 
finite-size scaling
exponents than the other known order parameters.
%%%%%%%
Importantly, we show that the conceptualization of the measures in terms of R\'{e}nyi and Tsallis 
entropies solves an 
%%uncomfortable 
incommodious
feature regarding the behavior of nearest-neighbor quantum discord in a second order phase transition.

There are at least two distinct ways in which the relative R\'{e}nyi and Tsallis entropies are defined, and are usually referred to as 
the ``traditional''~\cite{tradtsallis}
and ``sandwiched''~\cite{wilde,lennert} varieties. The sandwiched varieties incorporate the noncommutative nature of density matrices in an elegant way, and it 
is therefore natural to expect that it will play an important role in fundamentals and applications. Indeed, the sandwiched relative R\'{e}nyi entropy
%Recently, a quantum generalization of the R\'{e}nyi relative entropy has been introduced in \cite{wilde} 
%and~\cite{lennert} considering the non-commutativity 
%of a pair of density matrices. There has been a deluge of papers during the past few months which discuss several interesting properties
%of the generalized R\'{e}nyi relative entropy~\cite{wilde,lennert,beigi,lieb,rastegin}. As the generalized R\'{e}nyi relative entropy incorporate the non-commutate  nature of density
%matrices it is quite natural to expect that it is more usefull to interpret quantum phenomenon when non commutating density matrices are involved.
%In \cite{wilde} the generalized R\'{e}nyi relative entropy is properly coined as ``sandwiched'' 
has been used to show that the strong converse theorem for the classical capacity of a quantum channel holds for some specific channels~\cite{wilde}. 
Moreover, an operational
interpretation of the sandwiched relative R\'{e}nyi entropy in the strong converse problem of quantum hypothesis testing
 is noted for $\alpha>1$~\cite{ogawa}.
On the other hand, 
%in \cite{akr} follwing the work of \cite{wilde,lennert} 
the sandwiched relative Tsallis entropy  has recently been shown to be a better witness of entanglement \cite{akr}
than the traditional one \cite{abeakrqe}.
The relative min- and max-entropies \cite{rennerphd,renner1,nilans}, which can be obtained from the sandwiched relative R\'{e}nyi entropy for specific 
choices of $\alpha$,
play  significant roles in providing bounds on errors of one-shot entanglement 
cost~\cite{buscemi}, on the one-shot 
classical capacity of certain quantum channels~\cite{wang}, and in several scenarios in non-asymptotic quantum information theory~\cite{toma}.
In Ref. \cite{frustration}, connection of max- relative entopy with frustration in quantum many body systems has been established.
%quantum information 
 
% The corresponding relative min- and max-entropies 
% can be obtained from the sandwiched relative R\'{e}nyi entropy for specific choices of $\alpha$.
The paper is organized as follows.
In Sec.~\ref{gre}, we discuss the relative R\'{e}nyi and Tsallis entropies. In Sec.~\ref{QD}, we talk about the usual quantum discord. The 
R\'{e}nyi and Tsallis quantum correlations are defined in Sec.~\ref{gqd}, where we also derive their properties and evaluate them for paradigmatic
classes of bipartite quantum states. Some special cases like the ``linear'', ``min-'', and ``max-'' quantum discord are also 
formulated and discussed there. 
The quantum correlation measure is then applied for detecting quantum phase transition in a quantum many-body system in
Sec.~\ref{qpt}. We present a conclusion in Sec.~\ref{conclu}.%

\section{Relative R\'{e}nyi and Tsallis Entropies}
\label{gre}
The R\'{e}nyi \cite {renyi,renyiq} and Tsallis \cite{tsallis,tsallisq} entropies of a
density operator $\rho$ are given respectively by,\\

\begin{eqnarray}
\label{rt} 
S^{R}_\alpha(\rho)&=& \frac{1}{1-\alpha} \log \mbox{tr}[\rho^\alpha],  \\
 S^{T}_\alpha(\rho)&=& \frac{\mbox{tr}[\rho^\alpha]-1}{1-\alpha}.\\
 \end{eqnarray}

Here, the parameter $\alpha\in (0,1) \cup (1, \infty)$, unless mentioned otherwise. %for all quantum entropies and relative entropies.
 All logarithms in this paper are with base 2.
 Both the entropies reduce to the von Neumann entropy~\cite{von}, $S(\rho)= - \mbox{tr} (\rho \log \rho)$, when $\alpha\rightarrow 1$.
 In Ref. \cite{gen_q_entropy}, both the R\'{e}nyi and Tsallis entropies are derived from a generalized form of entropy and several 
 interesting properties of them are discussed.
The Tsallis entropy for $\alpha=2$ is called the 
 linear entropy, $S_L(\rho)$, given by  
 \begin{eqnarray}
  S_{L}(\rho)=1-\mbox{tr}[\rho^2].
 \end{eqnarray}
The traditional quantum relative R\'{e}nyi entropy between two density
operators $\rho$ and $\sigma$ is defined as
\begin{equation}
\label{rre}
  S_\alpha^{R}(\rho\vert\vert\sigma)=\dfrac{\log [\mbox{tr}\left( \rho^\alpha \sigma^{1-\alpha} \right)]}{\alpha-1}.
\end{equation}
Note that all the quantum relative entropies, traditional or sandwiched, discussed in this paper, are defined to be $+\infty$ if the kernel of $\sigma$
has non-trivial intersection with the support of $\rho$, and is finite otherwise.
$S_\alpha^{R}(\rho\vert\vert\sigma)$ reduces to the usual quantum relative entropy~\cite{umegaki}, $S(\rho||\sigma),$ when 
$\alpha\rightarrow 1$, where
\begin{eqnarray}
S(\rho||\sigma)=-S(\rho)-\mbox{tr}(\rho \log \sigma).
\end{eqnarray}
Recently, a generalized version of the quantum relative R\'{e}nyi entropy (called ``sandwiched'' 
relative R\'{e}nyi entropy) has been 
introduced, by considering the non commutative nature of density operators~\cite{wilde,lennert}. It is defined as
\begin{equation}
\label{srre}
\tilde{S}_\alpha^{R}(\rho\vert\vert\sigma)=
\dfrac{1}{\alpha-1}\log\left[\mbox{tr}\left(\sigma^{\frac{1-\alpha}{2\alpha}} 
\rho \sigma^{\frac{1-\alpha}{2\alpha}}\right)^\alpha\right].
% &\text{if 
% $\mbox{sp} (\rho) \subseteq \mbox{sp} (\sigma)$}.\\
%  \infty, & \mbox{otherwise}. \\
%  \end{cases}\\
%    \text{where $\alpha\in (0,1)\cup (1,\infty)$}. \nonumber 
\end{equation} 
Note that $\tilde{S}_\alpha^{R}(\rho\vert\vert\sigma)$ also reduces to $S(\rho||\sigma)$ when $\alpha\rightarrow 1$.
In Ref.~\cite{wilde,lennert,beigi,lieb,rastegin} several interesting properties of the sandwiched R\'{e}nyi entropy  
have been established. Here, we mention some of them (for two density operators
\(\rho\) and \(\sigma\)) which we will use later in this paper.
\begin{enumerate}
 \item 
 %For $\rho,\sigma\geq0$, $\rho\neq0$ and $\mbox{tr}(\rho)\geq\mbox{tr}(\sigma)$, 
 %we have 
 $\tilde{S}_\alpha^{R}(\rho\vert\vert\sigma)\geq0$.
 \item  $\tilde{S}_\alpha^{R}(\rho\vert\vert\sigma)=0$ if and only if $\rho=\sigma$.
 %for any  $\rho,\sigma\geq0$.
 \item For 
 %$\rho,\sigma\geq0$, $\rho\neq0$ and 
 $\alpha\in [\frac{1}{2},1) \cup (1, \infty)$ and for any completely positive 
 trace-preserving map $(\mbox{CPTPM})$ ${\cal E}$, we have the data processing inequality,
 $\tilde{S}_\alpha^{R}(\rho\vert\vert\sigma)\geq \tilde{S}_\alpha^{R}({\cal E}(\rho)\vert\vert{\cal E}(\sigma))$ \cite{lieb}. 
 %This relation, known as the data processing inequality, 
%  . In \cite{wilde,lennert} the data processing inequality was proved for $\alpha\in (1,2]$. 
%  It was recently shown in \cite{lieb} that data processing inequality actually 
 %holds for $\alpha\in [\frac{1}{2},1) \cup (1, \infty)$.%~\cite{wilde,lennert,lieb,beigi}.
%  and independently in \cite{beigi} for $\alpha>1$.
 \item $\tilde{S}_\alpha^{R}(\rho\vert\vert\sigma)$ is invariant under all unitaries $U$, i.e., 
 $\tilde{S}_\alpha^{R}(U\rho U^\dagger\vert\vert U\sigma U^\dagger) = \tilde{S}_\alpha^{R}(\rho\vert\vert\sigma)$ .
\end{enumerate}
%\vskip20pt

The traditional quantum relative Tsallis entropy between two density
operators $\rho$ and $\sigma$ is defined as
\begin{equation}
\label{rte}
S_\alpha^{T}(\rho\vert\vert\sigma)=\frac{ \mbox{tr}\left( \rho^\alpha \sigma^{1-\alpha} \right)-1}{\alpha-1}.
% \\
% & &\mbox{if}\  \mbox{supp} (\rho) \subseteq \mbox{supp} (\sigma).  \nonumber \\
% &&=\infty ,\mbox{otherwise}. \nonumber\\
% & & \mbox{where}\ \alpha \in (0,1) \cup (1, \infty) \nonumber
\end{equation}
% Following the work in \cite{wilde,lennert}, a quantum generalization of Tsallis entropy (called 
% ``sandwiched'' realtive Tsallis entropy) has been introduced in \cite{akr}.
The sandwiched relative Tsallis entropy between two density 
operators $\rho$ and $\sigma$ is given by~\cite{akr}
\begin{equation}
\label{srte}
\tilde{S}_\alpha^{T}(\rho\vert\vert\sigma)=\frac{\mbox{tr}\left[\left(\sigma^{\frac{1-\alpha}{2\alpha}} 
\rho \sigma^{\frac{1-\alpha}{2\alpha}}\right)^\alpha\right]-1}{\alpha-1}. 
% \\
% & &\mbox{if}\  \mbox{supp} (\rho) \subseteq \mbox{supp} (\sigma).  \nonumber \\
% &&=\infty ,\mbox{otherwise}. \nonumber\\
% & & \mbox{where}\  \alpha\in (0,1)\cup (1,\infty). \nonumber 
\end{equation} 
Both ${S}_\alpha^{T}(\rho\vert\vert\sigma)$ and $\tilde{S}_\alpha^{T}(\rho\vert\vert\sigma)$ also reduce to $S(\rho||\sigma)$ when $\alpha\rightarrow 1$.
 It can be easily verified that the properties (1-4), satisfied by $\tilde{S}_\alpha^{R}(\rho\vert\vert\sigma)$ 
 are also satisfied by $\tilde{S}_\alpha^{T}(\rho\vert\vert\sigma)$.
 In this paper, we will predominantly use the sandwiched version of both the relative entropies. Hereafter, by relative 
  entropy, we will  mean the sandwiched form 
 of the relative entropies, unless mentioned otherwise.
Some of the important special cases of the R\'{e}nyi and Tsallis relative entropies are given below.

\noindent 
{\bf a.} {\it Relative Linear Entropy}: At $\alpha=2$, $\tilde{S}_\alpha^{T}(\rho\vert\vert\sigma)$ gives the relative linear entropy,
\begin{equation}
\label{eijey}
 S_L(\rho\vert\vert\sigma)=\tilde{S}_2^{T}(\rho\vert\vert\sigma).
\end{equation}
The relative linear entropy has also been defined in the literature by using the traditional version of the 
relative entropy at \(\alpha = 2\). However, in this paper, we will use the relative linear entropy defined only through
the sandwiched relative entropy (at \(\alpha = 2\)).

\noindent {\bf b.} {\it Relative Collision Entropy}:  At $\alpha=2$, $\tilde{S}_\alpha^{R}(\rho\vert\vert\sigma)$ 
has been called the relative collision entropy \cite{rennerphd},
\begin{equation}
 S_C(\rho\vert\vert\sigma)=\tilde{S}_2^{R}(\rho\vert\vert\sigma).
\end{equation}
% Note that we are using the 
% %The relative linear entropy has also been defined in the literature by using the traditional version of the 
% %relative entropy at \(\alpha = 2\). However, in this paper, we will use the relative linear entropy defined only through
% %the 
% sandwiched relative entropy (at \(\alpha = 2\)) for defining the relative collision entropy.

\noindent {\bf c.} {\it Relative Min- and Max-Entropies}: In \cite{nilan}, 
it is pointed out that at $\alpha=\frac{1}{2}$, $\tilde{S}_\alpha^{R}(\rho\vert\vert\sigma)$
gives relative min-entropy~\cite{renner1},
\begin{equation}
 S_{min}(\rho\vert\vert\sigma)=\tilde{S}_{\frac{1}{2}}^{R}(\rho\vert\vert\sigma).
\end{equation}
Note that
\begin{eqnarray}
  S_{min}(\rho\Arrowvert\sigma)= -2\log F(\rho,\sigma),
 \end{eqnarray}
  $\mbox{where}\ F(\rho,\sigma)=\lVert\sqrt{\rho}\sqrt{\sigma}\rVert_1=\mbox{tr}\lvert\sqrt{\rho}\sqrt{\sigma}\rvert \nonumber$
  is the fidelity between the states \(\rho\) and \(\sigma\).
It is shown in \cite{lennert}, that the relative max-entropy \cite{nilans} is nothing but 
relative R\'{e}nyi entropy, when $\alpha\rightarrow\infty$ i.e.
\begin{equation}
S_{max}(\rho \Arrowvert \sigma)=\tilde{S}^R_{\alpha\rightarrow\infty}(\rho \Arrowvert \sigma), 
\end{equation}
where
\begin{eqnarray}
  S_{max}(\rho \Arrowvert \sigma)= \inf(\lambda : \rho \leq 2^{\lambda} \sigma).
  \end{eqnarray}

\section{Quantum discord}
\label{QD}

Quantum discord is a measure of quantum correlations of bipartite quantum states that is independent of the entanglement-separability 
paradigm \cite{hv,oz}. It can be conceptualized from several perspectives. An approach that is intuitively satisfying, is to 
define it as the difference between the total correlation and the classical correlation 
for a bipartite quantum state $\rho_{AB}$.
The total correlation is defined as the quantum mutual information of  \(\rho_{AB}\), which is given by 
\begin{equation}
\label{qmi}
\cal{I}(\rho_{AB})= S(\rho_A)+ S(\rho_B)- S(\rho_{AB}),
\end{equation}
where \(\rho_A\) and \(\rho_B\) are the local density matrices of \(\rho_{AB}\).
The mutual information $\cal{I}(\rho_{AB})$ can also be expressed in terms of the usual quantum relative entropy as
%  is also the relative entropy between $\rho_{AB}$ and the completely uncorrelated state $\rho_A\otimes \rho_B$, i.e.,
\begin{equation}
\label{mir}
 \cal{I}(\rho_{AB})=\min_{\{\sigma_A ,\sigma_B\}}S(\rho_{AB}||\sigma_A\otimes \sigma_B).
\end{equation}
This follows from the fact that 
\begin{eqnarray}
 %\lim_{\alpha\rightarrow1}\cal{I}_{\alpha}^{\varGamma}(\rho_{AB})&=&\lim_{\alpha\rightarrow1}\min_{\{\sigma_A ,\sigma_B\}}\tilde{S}^{\varGamma}_\alpha(\rho_{AB}\vert\vert\sigma_A\otimes \sigma_B).\nonumber\\
 &&\min_{\{\sigma_A ,\sigma_B\}}S(\rho_{AB}\vert\vert\sigma_A\otimes \sigma_B)\nonumber\\
 &=&\min_{\{\sigma_A ,\sigma_B\}}\{-S(\rho_{AB})-\mbox{tr}(\rho_A\log\sigma_A) \nonumber \\
 &-&\mbox{tr}(\rho_B\log\sigma_B)\},\nonumber
 \end{eqnarray}
and 
%Now from 
the non-negativity of relative von Neumann entropy between two density matrices.
% , we get
% \begin{eqnarray}
%   \lim_{\alpha\rightarrow1}\cal{I}_{\alpha}(\rho_{AB})=S(\rho_{AB}\vert\vert\rho_A\otimes \rho_B)=\cal{I}^{}(\rho_{AB}).
% \end{eqnarray}
% 
% 
% 
Therefore, the quantum mutual information is the minimum usual relative entropy distance of the state $\rho_{AB}$ from the set of all completely uncorrelated states,
$\sigma_A\otimes \sigma_B$, whence we obtain a ground for interpreting the quantum mutual information as the total correlation in the state. Further
evidence in this direction is provided in \cite{qmi,Cerf,GROIS,Barry}.
The classical correlation is given in terms of the measured conditional entropy, and is defined as \cite{hv,oz}
\begin{equation}
\label{cc}
 \cal{J}(\rho_{AB}) = S(\rho_A) - S(\rho_{A|B}), 
\end{equation}
where
\begin{equation}
\label{ccd}
 S(\rho_{A|B}) = \min_{\{P_i\}} \sum_i p_i S(\rho_{A|i})
\end{equation}
is the conditional entropy of \(\rho_{AB}\), conditioned on measurements at \(B\) with rank-one 
projection-valued measurements \(\{P_i\}\).
 Here, \(\rho_{A|i} = \frac{1}{p_i} \mbox{tr}_B[(\mathbb{I}_A \otimes P_i) \rho (\mathbb{I}_A \otimes P_i)]\) is the conditional state which we 
 get 
with probability \(p_i = \mbox{tr}_{AB}[(\mathbb{I}_A \otimes P_i) \rho (\mathbb{I}_A \otimes P_i)]\), where
\(\mathbb{I}_A\) is the identity operator on the Hilbert space of \(A\). $\cal{J}(\rho_{AB})$ 
can also be defined in terms of the mutual information as
\begin{equation}
\label{ccr}
 \cal{J}(\rho_{AB})=\max_{\{P_i\}} \cal{I}(\rho'_{AB}),
\end{equation}
where
\begin{equation}
\label{ro'}
\rho'_{AB}=\sum_i(\mathbb{I}_A \otimes P_i) \rho_{AB} (\mathbb{I}_A \otimes P_i).
\end{equation}
The classical correlation can therefore be seen as the minimum relative entropy distance of the state $\rho'_{AB}$ from all uncorrelated states,
 maximized over all rank-one projective measurements on $B$, and is given by
 \begin{equation}
 \label{jro}
   \cal{J}(\rho_{AB})=\max_{\{P_i\}}\min_{\{\sigma_A ,\sigma_B\}}S(\rho'_{AB}||\sigma_A\otimes \sigma_B).
 \end{equation}
 The maximization in Eq. (\ref{jro}) or in Eq. (\ref{cc}) ensure that ${\cal J}(\rho_{AB})$ quantifies the 
 maximal content of classical correlation present in the bipartite state $\rho_{AB}$.
 Hence, if we subtract $\cal{J}(\rho_{AB})$ from the total correlation, the remaining
 correlation is ``purely'' quantum, and is defined as \cite{hv,oz}
\begin{equation}
\label{diseqn}
{\cal D}(\rho_{AB})= {\cal I}(\rho_{AB}) - {\cal J}(\rho_{AB}).
\end{equation}

\section{Total, Classical, and Quantum Correlations as Relative Entropies}
\label{gqd}
In this section, we define the total, classical, and quantum correlation in terms of the sandwiched relative R\'{e}nyi and Tsallis entropies. We discuss
the properties of these measures and evaluate them for several important families of bipartite quantum states. In the final subsection, we also compare the results
with those obtained with traditional relative entropies.
% should satisfy some necessary properties. In this section we discuss the properties and then show that our measure of quantum correlation satisfies all the necessary properties.
\subsection{Generalized Mutual Information as Total Correlation}
 We define the generalized mutual information of  $\rho_{AB}$ as
 \begin{eqnarray}
 \label{rmi}
   \cal{I}^{\varGamma}_\alpha(\rho_{AB})=\min_{\{\sigma_A ,\sigma_B\}}\tilde{S}_\alpha^{\varGamma}(\rho_{AB}\vert\vert\sigma_A\otimes \sigma_B).
 \end{eqnarray}
%where the relative entropy is defined either through  
Here, the minimum is taken over all density matrices, $\sigma_A$ and $\sigma_B$. The relative entropy, although 
 not a metric on the operator space, is a measure of the distance between two quantum states.
  $\tilde{S}_\alpha^{\varGamma}(\rho_{AB}\vert\vert\sigma_A\otimes \sigma_B)$ is a
 distance between the quantum state $\rho_{AB}$ and a completely uncorrelated state $\sigma_A\otimes \sigma_B$. Here, and hereafter, the superscript
 $\varGamma$ is either $R$ or $T$, depending on whether it is the R\'{e}nyi or Tsallis variety that is considered.
%  Now if we take the infimum of the distances of $\rho_{AB}$ from all
%   kind of $\sigma_A\otimes \sigma_B$ then it can be considered as total correlation among the two parties of $\rho_{AB}$ \cite{vedre}. 
 The corresponding minimum distance can be interpreted as the total correlation present in the system.
The generalized  mutual information $ \cal{I}^{\varGamma}_\alpha(\rho_{AB})$
 becomes equal to the usual quantum mutual information $\cal{I}(\rho_{AB})$ when $\alpha\rightarrow1$: 
 \begin{eqnarray}
 \lim_{\alpha\rightarrow1}\cal{I}_{\alpha}^{\varGamma}(\rho_{AB})&=&\lim_{\alpha\rightarrow1}\min_{\{\sigma_A ,\sigma_B\}}\tilde{S}^{\varGamma}_\alpha(\rho_{AB}\vert\vert\sigma_A\otimes \sigma_B).\nonumber\\
 &=&\min_{\{\sigma_A ,\sigma_B\}}S(\rho_{AB}\vert\vert\sigma_A\otimes \sigma_B)\nonumber\\
 &\equiv& \cal{I}^{}(\rho_{AB}).
 %\min_{\{\sigma_A ,\sigma_B\}}\{-S(\rho_{AB})-\mbox{tr}(\rho_A\log\sigma_A) \nonumber \\
 %&-&\mbox{tr}(\rho_B\log\sigma_B)\}.\nonumber
 \end{eqnarray}
% Now from the positivity of relative von Neumann entropy between two density matrices, we get
% \begin{eqnarray}
%   \lim_{\alpha\rightarrow1}\cal{I}_{\alpha}(\rho_{AB})=S(\rho_{AB}\vert\vert\rho_A\otimes \rho_B)=\cal{I}^{}(\rho_{AB}).
% \end{eqnarray}

%  In similar manner Tsallis mutual information is defined by just replacing the 
%  the relative R\'{e}nyi Entropy by relative Tsallis entropy in (\ref{rmi}).

% \subsection{Necessary Properties of Quantum Correlation Measure} 
% \label{propqc}
% While defining the quantum correlation measure like quantum discord where the difference between total and classical correlation is considered to be
% quantum correlation we have to be careful as the quantum correlation measure is expected to satisfy the following properties :-
% \begin{enumerate}
%  \item $\cal{TC}, \cal{CC} \geq0$.
%  \item $\cal{TC}=0$ for $\rho_{AB}=\rho_A\otimes \rho_B$.
%  
%  \item $\cal{TC}$ is invariant under local unitaries.
%  \item $\cal{TC}$ is non-increasing under local operations.
%  \item$\cal{TC} \geq \cal{CC}$, so that $\cal{QC} \geq 0$.
%  \item Measure of $\mbox{CC}$ is expected to satisfy the following properties given in \cite{hv} :-
%  \begin{enumerate}
%   \item $\cal{CC}=0$ for $\rho_{AB}=\rho_A\otimes \rho_B$.
%   \item $\cal{CC}$ is invariant under local unitaries.
%   \item $\cal{CC}$ is non-increasing under local operations.
%  \end{enumerate}
% 
% \end{enumerate}
% These properties are necessary to satisfy for a `good' quantum correlation measure.

 \subsection{Classical and Quantum Correlation}
 \label{qdr}
%  We now define the quantum discord in terms of the R\'{e}nyi entropy which satisfies all the properties discussed in the previous subsection.
   The R\'{e}nyi or Tsallis version of the classical correlation,  
 denoted by $\cal{J}^{\varGamma}_\alpha(\rho_{AB})$,  is defined as
 \begin{eqnarray}
 \label{gcc}
  \cal{J}^{\varGamma}_\alpha(\rho_{AB})=\max_{\{P_i\}}\min_{\{\sigma_A ,\sigma_B\}}\tilde{S}_\alpha^{\varGamma}(\rho'_{AB}\vert\vert\sigma_A\otimes \sigma_B),
 \end{eqnarray}
where $\rho'_{AB}$ is obtained by performing rank-1 projective measurements as 
in the definition  of original classical correlation (in Eq. (\ref{ro'})).
% These definition of the total correlation and classical correlation satify
%  all the properties mentioned in (\ref{propqc}). 

Therefore, quantum correlation using generalized entropies is defined as
 \begin{equation}
 \cal{D}^{\varGamma}_\alpha(\rho_{AB})= \cal{I}^{\varGamma}_\alpha(\rho_{AB})-\cal{J}^{\varGamma}_\alpha(\rho_{AB}),
 \end{equation}
 with $\alpha\in [\frac{1}{2},1)\cup (1,\infty)$. By using the data processing inequality, which holds in this range of $\alpha$,
 one can prove the non-negativity of the quantum correlation~\cite{lieb}. 
 We now look into the properties of $\cal{D}^{\varGamma}_\alpha(\rho_{AB})$, which provide independent support for 
 identifying the quantities as correlation measures.
 %\vskip10pt
 
 \noindent{\bf Property 1}. $\cal{I}^{\varGamma}_\alpha, \cal{J}^{\varGamma}_\alpha \geq0$ 
 since $\tilde{S}_\alpha^{\varGamma}(\rho\vert\vert\sigma)\geq0$.
 
\noindent{\bf Property 2}. $\cal{I}^{\varGamma}_\alpha, \cal{J}^{\varGamma}_\alpha$ are 
vanishing, and therefore, $\cal{D}^{\varGamma}_\alpha = 0$, 
for any product state, $\rho_{AB}=\rho_A\otimes \rho_B$, 
 as $\tilde{S}_\alpha^{R}(\rho\vert\vert\rho)=0$. The proof for the vanishing of total correlations follows by noting that 
 the product state in the argument itself is the state which gives the optimal relative entropy distance. A similar 
 argument, but for the measured state, holds for 
 the classical correlation.\\
 Moreover, $\cal{D}^{\varGamma}_\alpha = 0$ for any quantum-classical state, i.e. any state of the form 
 $\sum_i p_i\rho_i^A\otimes(|i\rangle\langle i|)^B$,
 where $\{p_i\}$ forms a probability distribution, $\{|i\rangle\}$ forms an orthonormal basis, and $\rho_i$ are density matrices, 
 when the measurement is performed on the $B$ part.
 
 \noindent{\bf Property 3}. $\cal{I}^{\varGamma}_\alpha, \cal{J}^{\varGamma}_\alpha$ remain invariant under local unitaries, which follow from the fact that 
 $\tilde{S}_\alpha^{R}(\rho\vert\vert\sigma)$ is invariant under all unitaries $U$. Hence, $\cal{D}^{\varGamma}_\alpha$ is also invariant under local
 unitaries.
 
 \noindent{\bf Property 4}. $\cal{I}^{\varGamma}_\alpha, \cal{J}^{\varGamma}_\alpha$ are non increasing under local operations, which follow from the data processing inequality,
 $\tilde{S}_\alpha^{R}(\rho\vert\vert\sigma)\geq \tilde{S}_\alpha^{R}({\cal E}(\rho)\vert\vert{\cal E}(\sigma))$, for any $\mbox{CPTPM}$ ${\cal E}$.
 
 \noindent{\bf Property 5}. $\cal{D}^{\varGamma}_\alpha$ is non-negative, as $\cal{J}^{\varGamma}_\alpha$ is upper bounded 
 by $\cal{I}^{\varGamma}_\alpha$. The latter statement is due to the fact that $\cal{J}^{\varGamma}_\alpha$ is obtained 
 %by 
 %from $\cal{I}^{\varGamma}_\alpha$ 
 by performing 
 a local measurement on \(\rho_{AB}\), and we know from the 
% which 
% again follows from the 
data processing inequality that \(\tilde{S}_\alpha^\varGamma\) is monotone under CPTPM. 
 %\vskip10pt
 
 The classical correlation measure that 
we have defined here, satisfies all the plausible properties for classical correlation proposed in Ref. \cite{hv}, except the one which states that 
for pure states, 
the classical correlation reduces to the von Neumann
entropy of the subsystems.
%In this case,  
We wish to mention that this property is natural for the measure which involves 
%very much particular to 
the von Neumann entropy, and is not expected to be followed by the 
measures with generalized entropies. This is because the definition of classical correlation in terms of the relative entropy reduces naturally to the 
one in terms of the conditional entropy in the case of the von Neumann entropy. 

We use the convention that each of the definitions of $\cal{I}^{\varGamma}_\alpha$,  $\cal{J}^{\varGamma}_\alpha$ and $\cal{D}^{\varGamma}_\alpha$ also incorporates
a division by $\log{2}$ bits, whence all the definitions can be considered to be dimensionless.

We note here that there has been previous attempts to define quantum discord by using Tsallis entropies~\cite{tsallisdis1,tsallisdis2,tsallisdis3}.
These definitions however do not always guarantee positivity of the quantum discord, so defined. Also, the 
corresponding total and classical correlations are not
necessarily monotonic under local operations. Ref. \cite{tsallisdis4} defines a quantum correlation by considering the 
difference between the Tsallis entropies of the post-measured and pre-measured states. 
% Further works in Refs. \cite{tsallisdis4} and \cite{renyidis,wildemutual}
% define quantum correlations by using Tsallis and R\'{e}nyi 
% entropies respectively, whose approaches however are different from ours.
In Ref. \cite{renyidis}, a Gaussian quantum correlation is defined by using the R\'{e}nyi entropy for $\alpha=2$.
After completion of the current paper, we came to know about the work in Ref. \cite{wildemutual}, which
states that  quantum discord can be defined using  sandwiched relative entropy, with a  definition 
of quantum mutual information of a bipartitite quantum state $\rho_{AB}$, given by
$\cal{I'}^{R}_\alpha(\rho_{AB})=\min_{\{\sigma_B\}}\tilde{S}_\alpha^{R}(\rho_{AB}\vert\vert\rho_A\otimes \sigma_B)$ (cf. Eq. (\ref{rmi})). 
The definition of mutual information used here and in Ref. \cite{ourpre} (published version of this paper) use \(\sigma_A\) in place of \(\rho_A\), which makes \(\cal{I'}^{R}_\alpha\),
a special case of it. By ``special case'', it is meant that the optimization performed in this work is over a class of states that contains the class of states used in \cite{wildemutual}. The R\'enyi
quantum discord was later defined by the authors of \cite{wildemutual} in terms of generalized conditional
mutual information \cite{wildediscord}, an approach that is very different from the one
followed in this work.
% \textcolor{red}{In \cite{ourpre} (published version of the current paper), we have stated 
% that ``the definition of mutual information in the current paper is more general, since we use \(\sigma_A\) in place of \(\rho_A\), which makes \(\cal{I'}^{R}_\alpha\),
% a special case
% of the mutual information defined in this paper \cite{wildepub} (which is the published version of \cite{wildemutual}).'' By ``special case'', we  mean that the optimization performed in this paper is over a class of states that contains the class of states used in \cite{wildemutual} or \cite{wildepub}. Later, the authors of \cite{wildemutual} and \cite{wildepub} defined R\'{e}nyi quantum discord in terms of generalized conditional mutual information \cite{wildediscord} and the approach is very much different from the ones followed in this work.}
% with 
% different approaches from us.

 \subsection{Special Cases}
\subsubsection{ Linear Quantum Discord}

The relative linear entropy can be used to define the  ``linear quantum discord'', given by 
\begin{equation}
  \cal{D}_L(\rho_{AB})= \cal{I}^T_2(\rho_{AB})-\cal{J}^T_2(\rho_{AB}),
\end{equation}
where \(\cal{I}^T_2(\rho_{AB})\) and \(\cal{J}^T_2(\rho_{AB})\) are defined by using the relative linear entropy, given in Eq. (\ref{eijey}).
\subsubsection{ Min- and Max-Quantum Discords}
We also define the ``min- and max-quantum discords''  by considering relative min- and max-entropies as
\begin{equation}
 \cal{D}_{min}(\rho_{AB})= \cal{I}^R_{\frac{1}{2}}(\rho_{AB})-\cal{J}^R_{\frac{1}{2}}(\rho_{AB}),
\end{equation}
 and 
 %max- quantum discord is given by
\begin{equation}
 \cal{D}_{max}(\rho_{AB})= \cal{I}^R_{\alpha\rightarrow\infty}(\rho_{AB})-\cal{J}^R_{\alpha\rightarrow\infty}(\rho_{AB}). 
\end{equation}

 \subsection{Pure States}
 \label{eijey!!}
  Any bipartite pure state of two qubits can be written, using Schmidt decomposition, as
%   As generalized quantum discord is local unitarilly invariant like the usual quantum discord to calculate generalized 
%   quantum discord for a bipartite   pure we can write the state in Schimdt basis
%   without loss of generality. So we write an arbitrary bipartite pure state as 
\begin{equation}
 |\psi_{AB}\rangle=\displaystyle\sum_{i=0}^1 \sqrt{\lambda_i} |i_{A}i_{B}\rangle,
\end{equation}
where $\lambda_i$ are non-negative real numbers satisfying $\sum_i\lambda_i=1$.
%   Note that the state is symmetric with respect to the parties $A$ and
%   $B$ i.e. we can make any bipartite pure state symmetric by choosing proper basis. Now we have to calculate the 
%   total as well as classical correlation for this state. \\
%   Total Correlation\\
  Since a bipartite pure state is symmetric, it is expected that the state $\sigma_A\otimes\sigma_B$, which minimizes the relative 
  entropy of $|\psi_{AB}\rangle$ with uncorrelated states, 
%   we name it as  $\sigma_A^{min}\otimes \sigma_B^{min}$, 
  is also symmetric. Numerical studies support this view. This fact is not only true 
  for pure bipartite states, but it holds for all symmetric
  bipartitite states that are considered in this paper.
  %
%   The justification 
%   for this conjecture is that as $\vert\psi\rangle$ is symmetric in the parties $A$ and $B$ there should not be any 
%   way to distinguish $A$ and $B$. But if $\sigma_A^{min}\otimes \sigma_B^{min}$
%   is not symmetric then there is a way to distinguish between $A$ and $B$. So we argue that for any symmetric 
%   $\rho_{AB}$ the $\sigma_A^{min}\otimes \sigma_B^{min}$ 
%   is also symmetric i.e. $\sigma_A^{min}\otimes \sigma_B^{min}= \sigma^{min}\otimes \sigma^{min}$. It seems quite 
%   legitimate too. Nonetheless there 
%   is still another possibility. Suppose there is one $\sigma_A^{min}\otimes \sigma_B^{min}$ which is not symmetric
%   but still we get the minimum relative entropy for it,
%   then invoking the symmetry arguement we can say there is another 
%   $\sigma_B^{min}\otimes \sigma_A^{min}$ which also minimizes relative entropy. But here with the support of
%   numerical studies we conjecture that we can achieve the minimum for the symmetric state
%   i.e., $\sigma_A^{min}\otimes \sigma_B^{min}= \sigma^{min}\otimes \sigma^{min}$.\\
  Moreover, numerical results indicate that for arbitrary $\vert\psi_{AB}\rangle$, the state $\sigma^{A}\otimes \sigma^{B}$ which gives the minimum,
  is diagonal in the Schmidt basis of $\vert\psi_{AB}\rangle$.
 To numerically evaluate the minimum relative entropy distance of a bipartite quantum state $\rho_{AB}$ from  product states, we begin by 
  randomly 
  generating bipartite product states $\sigma_A\otimes\sigma_B$. Then we calculate the relative entropies between $\rho_{AB}$ and all 
  such $\sigma_A\otimes\sigma_B$. The minimum of these relative entropies is considered to 
  be the minimum relative entropy distance. We repeat the procedure for a larger set of randomly chosen product states. We terminate the process
  %halt the numerical procedure 
  when the minimum does not change within the required precision. Note that the numerical study is performed without the assumptions that 
  the product state at which the minimum is attained is symmetric and that it is diagonal in the Schmidt basis. 
  We have followed the same procedure throughout the paper
  to numerically evaluate the different correlations.
 Therefore, the minimum $\sigma_{A}$ or $\sigma_{B}$ is given by
  \begin{equation}
   \sigma_{A}=\sigma_{B}=\sigma\equiv\displaystyle\sum_{i=0}^1 a_i\vert i\rangle\langle i\vert,
  \end{equation}
where $a_i$ are non-negative real numbers satisfying $\sum_ia_i=1$.
% $\sigma^{min}=p\vert0\rangle\langle0\vert+(1-p)\vert1\rangle\langle1\vert$. Where \(0\leq p \leq 1\).
  With these assumptions, the total correlation
  %relative R\'{e}nyi entropy 
  of $|\psi_{AB}\rangle$ 
  %with 
 %$\sigma\otimes \sigma$, 
  is given by 
\begin{eqnarray}
\label{puretc}
  \cal{I}^R_\alpha(\vert\psi_{AB}\rangle)&=&\min_{\{a\}}\frac{1}{\alpha-1} \log\Big[\lambda a^{\frac{2(1-\alpha)}{\alpha}} \nonumber \\
  &+&  (1-\lambda)(1-a)^{\frac{2(1-\alpha)}{\alpha}}\Big]^\alpha, 
\end{eqnarray}
where $a_0=a$, $a_1=1-a$, $\lambda_0=\lambda$, $\lambda_1=1-\lambda$.
The value of $a$ is obtained from the condition
%  Now we have to find the value of $p$ in terms of $\alpha$ and $a$ from the minimization condition. 
%  From the minimization condition, for $\alpha\geq1$ we have
 \begin{equation}
  \frac{1}{a}=\left(\frac{\lambda}{1-\lambda}\right)^{\frac{\alpha}{2-3\alpha}} +1,
 \end{equation}
 for $\alpha\in (2/3,1) \cup (1, \infty)$. For $\frac{1}{2}\leq\alpha\leq\frac{2}{3}$, the minimization in Eq. (\ref{puretc}) yields
 \begin{equation}
  \cal{I}^R_\alpha(\vert\psi_{AB}\rangle)=\frac{\alpha}{\alpha-1} \log\big[\max\{\lambda,1-\lambda\}\big]. 
\end{equation}
%  For $\alpha<1$,  $a p^{\frac{2(1-\alpha)}{\alpha}}+(1-a)(1-p)^{\frac{2(1-\alpha)}{\alpha}}$ is less than one 
%  (as relative entropy is always non-negative) and to 
%  minimize the relative entropy we have to maximize $a p^{\frac{2(1-\alpha)}{\alpha}}+(1-a)(1-p)^{\frac{2(1-\alpha)}{\alpha}},
%  (=X say)$. One can easily see that for
%  $\frac{2}{3}\leq\alpha<1$, $X$ is maximized for \\
%  $ \frac{1}{p}=(a/(1-a))^{\frac{\alpha}{2-3\alpha}} +1$. So the relative entropy is minimum at this point for 
%  $\alpha\geq\frac{2}{3}$. But for $\frac{1}{2}\leq\alpha<\frac{2}{3}$
%  $X$ is minimized at this point and dont have any maxima. 
%  
%  Classical Correlation.\\
  For pure states, numerical searches indicate that the classical correlation is independent of the measurement basis.
  We consider measurement performed in the Schmidt basis for 
 calculating the classical correlation of the original state.
 Just like the total correlation in the original state, the  $\sigma_A\otimes\sigma_B$, which minimizes the relative 
  entropy of the post-measurement state with uncorrelated states, is symmetric, since we perform the 
  projective measurement in the Schmidt basis. Moreover, from numerical 
  results, we find that $\sigma_A\otimes\sigma_B$ is again diagonal in the Schmidt basis of $\vert\psi_{AB}\rangle$.
%  i.e., we
%  have to calculate the total correlation of the mixed state 
%  $a\vert00\rangle\langle00\vert+(1-a)\vert11\rangle\langle11\vert$ to find the classical correlation of $\vert\psi\rangle$. 
%  Here also invoking the symmetry arguement and considering numerical evidences
%  we take $\sigma_A^{min}\otimes \sigma_B^{min}= \sigma^{min}\otimes \sigma^{min}$. Numerically we have also seen that
%  $\sigma^{min}$ is diagonalised in the Schimdt basis of $\psi$.
%  So we take $\sigma^{min}=q\vert0\rangle\langle0\vert+(1-q)\vert1\rangle\langle1\vert$. 
The R\'{e}nyi classical correlation of $\vert\psi_{AB}\rangle$ is therefore given by
 \begin{eqnarray}
  \cal{J}^R_\alpha(\vert\psi_{AB}\rangle)&=&\min_{\{a\}}\frac{1}{\alpha-1} \log\Big[ \lambda^{\alpha} a^{2(1-\alpha)} \nonumber \\
  &+&(1-\lambda)^{\alpha}(1-a)^{2(1-\alpha)}\Big]. 
\end{eqnarray}
The value of $a$ is obtained from the condition
% From the minimization condition we get  
\begin{equation}
  \frac{1}{a}=\left(\frac{\lambda}{1-\lambda}\right)^{\frac{\alpha}{1-2\alpha}} +1,
 \end{equation}
 for $\alpha\in (1/2,1) \cup (1, \infty)$.
 %\vskip10pt
 
 The linear quantum discord for $\vert\psi_{AB}\rangle$ is given by
 \begin{equation}
 \cal{D}_{L}(\vert\psi_{AB}\rangle)=\big(\sqrt{\lambda}+\sqrt{1-\lambda} \big)^4-\big(\sqrt{\lambda}+\sqrt{1-\lambda}\big)^2.
\end{equation}

 We find that the min-quantum discord is vanishing for every two-qubit pure state.
 We believe that this is a peculiarity of some elements of the class of information-theoretic 
 quantum correlation measures that are defined according to the premise that subtracting 
 classical correlations from total correlations will produce quantum correlations. 
 This may perhaps be paralleled with the fact that although it 
 was perhaps considered desirable that all entanglement measures should possess the property that they 
 should vanish for separable states and only for separable states, the discovery of bound entangled
 states \cite{bound} led us to the fact that distillable entanglement \cite{dist_entngl} can vanish for certain entangled states as well.
It should be noted that in contradistinction to distillable entanglement, the min-quantum discord 
can be non-zero for certain separable states, indicating that at least in this sense, the space of 
information-theoretic quantum correlations is richer than the space of entanglement measures.
 \begin{figure}[htb]
\begin{center}
   \includegraphics[scale=0.5]{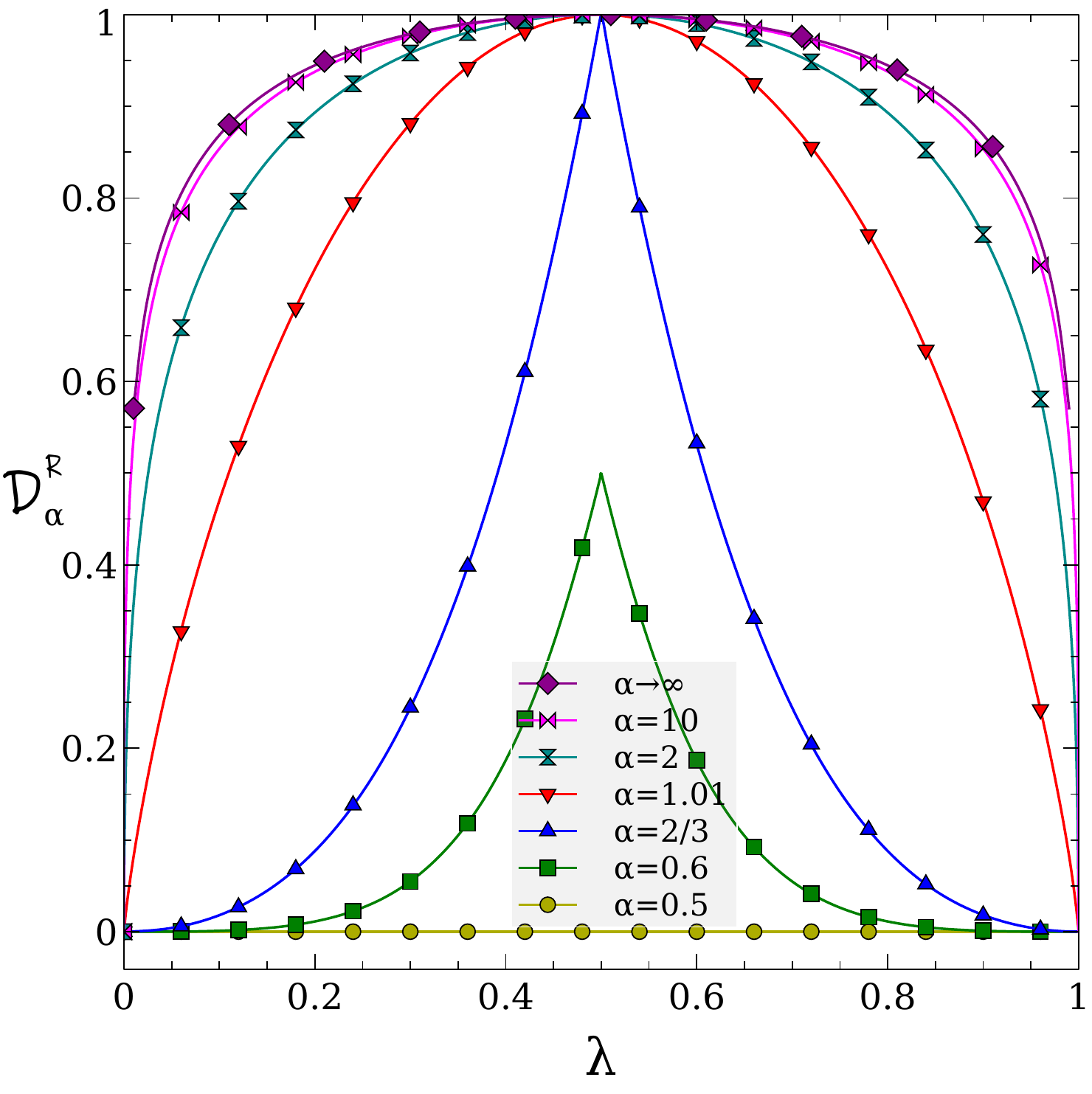}
\end{center}
\caption{(Color online.) R\'{e}nyi quantum correlation, $\cal{D}^R_\alpha$, with respect to $\lambda$, of $\vert\psi_{AB}\rangle=
\sqrt{\lambda}|00\rangle+\sqrt{(1-\lambda)}|11\rangle$,  for different $\alpha$. Both axes are dimensionless. }
 \label{fig-pure}
 \end{figure}
 
 The max-quantum discord for $\vert\psi_{AB}\rangle$ is given by
\begin{equation}
 \cal{D}_{max}(\vert\psi_{AB}\rangle)=\log\left[\frac{(\sqrt[3]{\lambda}+\sqrt[3]{1-\lambda})^3}{(\sqrt{\lambda}+\sqrt{1-\lambda})^2}\right].
\end{equation}

\begin{figure}[htb]
\begin{center}
   \includegraphics[scale=0.5]{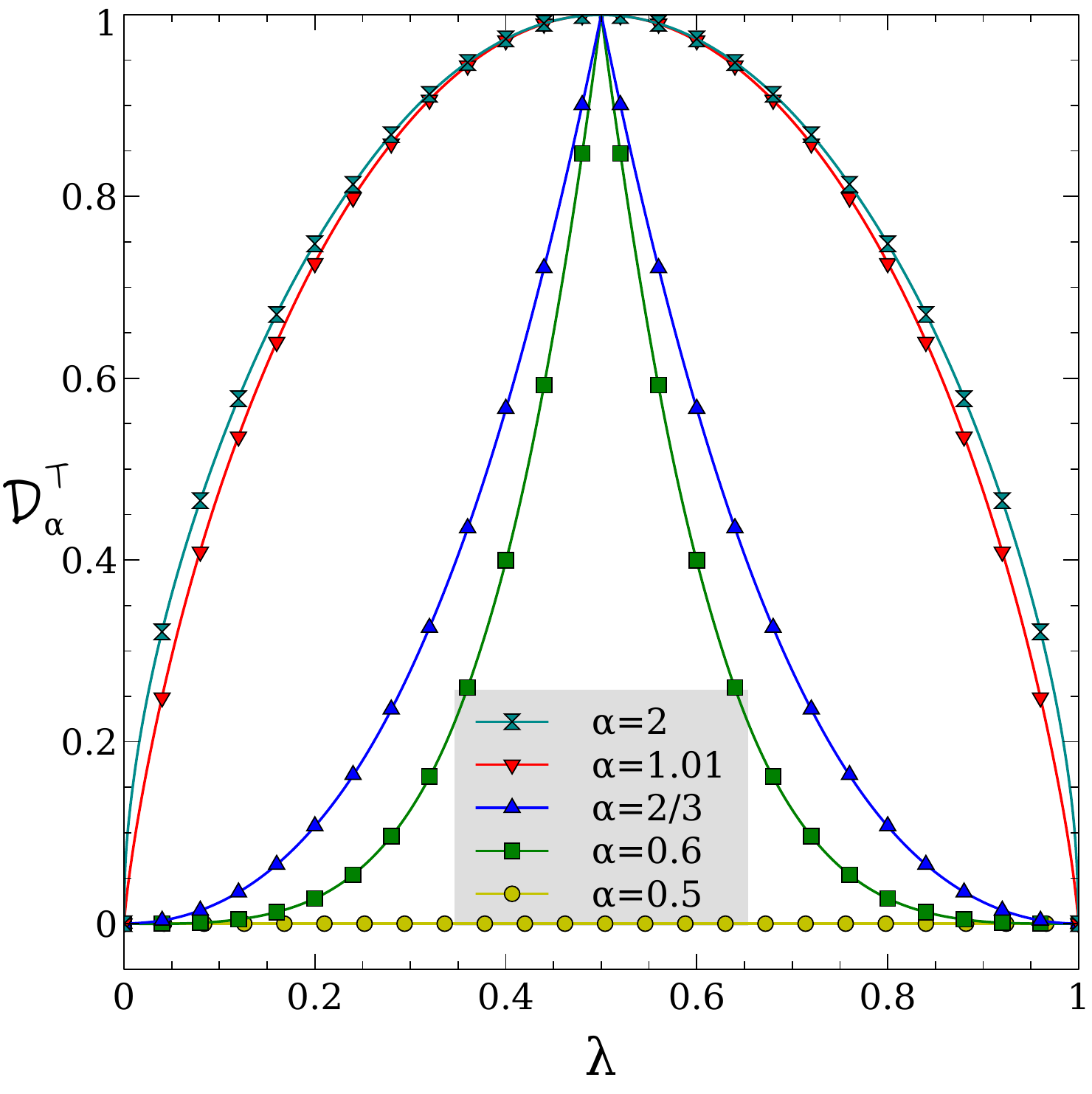}
\end{center}
\caption{(Color online.) Tsallis quantum correlation, $\cal{D}^T_\alpha$, with respect to $\lambda$, of $\vert\psi_{AB}\rangle=
\sqrt{\lambda}|00\rangle+\sqrt{(1-\lambda)}|11\rangle$,  for different $\alpha$. Both axes are dimensionless. The values of the Tsallis quantum correlation are 
normalized, whenever possible, so that the maximal quantum correlations are of unit value.}
 \label{fig-pure-tsallis}
 \end{figure}

 In Fig.~\ref{fig-pure}, we plot the R\'{e}nyi quantum correlation of $\vert\psi_{AB}\rangle$ for various values of $\alpha$. 
 We have also performed the entire calculations
 for the Tsallis discord and find that its behavior is qualitatively  similar to the R\'{e}nyi discord. In Fig.~\ref{fig-pure-tsallis},
 we have exhibited the Tsallis discord for bipartite pure states, which clearly indicate  the similarity
 between the two discords. In the rest of the paper, we will only plot the R\'{e}nyi discord.

%  And the minima is achieved for $\alpha\geq\frac{1}{2}$. 
 \subsection{Mixed States: Some Examples}
 \label{renyixam}
%   Here we investigate behavior of the R\'{e}nyi quantum discord for some examples.\\
\noindent {\bf (i) Werner States:} Consider the Werner state, given by
  \begin{equation}
\rho_{W}=p|\psi^-\rangle\langle\psi^-| +(1-p)\frac{I}{4}, \nonumber
\end{equation}
 where $|\psi^-\rangle=\frac{1}{\sqrt{2}}(|01\rangle-|10\rangle)$, \(I\) denotes the 
identity operator on the two-qubit Hilbert space, and \(0\leq p \leq 1\).
%  First we want to find the total correlation of werner state for different values of $\alpha$. So we have to find 
%  the optimal $\sigma_A$ and $\sigma_B$.  
 Suppose the $\sigma_A^{min}$ and $\sigma_B^{min}$ are the optimal $\sigma_A$ and $\sigma_B$ which minimizes the 
 relative R\'{e}nyi entropy of $\rho_{W}$ with  uncorrelated states.
%  $\sigma_A^{min}$ and $\sigma_B^{min}$ are diagonalised in any arbitrary basis and a priori we dont know the 
%  basis. Whatever may be the basis we can name it as $\vert0\rangle$, 
%  $\vert1\rangle$ for both $\sigma_A^{min}$ and $\sigma_B^{min}$. So 
Using the fact that the Werner state is symmetric 
%\textcolor{red}{(as mentioned in Sec. \ref{eijey!!})} 
and local unitarily invariant, we  choose 
% $\sigma_A^{min}\otimes \sigma_B^{min}$ can be written as
%  given by
 \begin{eqnarray}
  \sigma_A^{min}= \sigma_B^{min}=\sigma\equiv\displaystyle a_0\vert 0\rangle\langle 0\vert + 
  a_1\vert 1\rangle\langle 1\vert,
  %&\otimes&(b\vert0\rangle\langle0\vert+  (1-b)\vert1\rangle\langle1\vert),\nonumber 
  \end{eqnarray}
 where $a_i$ are non-negative real numbers satisfying $\sum_ia_i=1$.
Here we have assumed that $\sigma_A\otimes\sigma_B$, which minimizes the relative 
  entropy of $\rho_{W}$ with uncorrelated states, is symmetric. Detail numerical study support our assumption, as mentioned in Sec. \ref{eijey!!}.
%   we name it as  $\sigma_A^{min}\otimes \sigma_B^{min}$, 
  %is also symmetric
%  Now we have to find the value of $a$ and $b$ to find $\sigma_A^{min}\otimes \sigma_B^{min}$. We know
%  that singlet is local unitarily invariant.
It is now possible to perform the minimization 
%is now possible We find
% . We can write the werner state in the same basis in which
%  $\sigma_A^{min}\otimes \sigma_B^{min}$ is diagonalised. To calculate the relative entropy
%  we have to calculate trace of the matrix
%  \begin{equation}
%   X=((\sigma_A^{min}\otimes \sigma_B^{min})^{\frac{1-\alpha}{2\alpha}} \rho (\sigma_A^{min}\otimes 
%   \sigma_B^{min})^{\frac{1-\alpha}{2\alpha}})^\alpha \nonumber
%  \end{equation}
% 
%  We have analytically found  
% that 
for $\alpha \in [\frac{2}{3},1) \cup (1,\infty)$. In this range, the relative R\'{e}nyi entropy distance corresponding to the 
total correlations
%  trace of $X$ 
 is minimum for $a_0=a_1=\frac{1}{2}$. 
%  i.e. the minima of relative R\'{e}nyi entropy is achieved for \\
%  \begin{equation}
%  \sigma_A^{min}\otimes \sigma_B^{min}=\frac{I}{4}.\nonumber
%  \end{equation}
%  
%  For $\alpha<1$ we have to maximize the 
%  trace of $X$ to achieve the minima of the R\'{e}nyi relative entropy. We have found analytically that for 
%  $\frac{2}{3}\leq\alpha<1$ the trace maximizes at $a=\frac{1}{2}$, $b=\frac{1}{2}$. So in this range 
%  the R\'{e}nyi relative entropy also minimizes for 
%   \begin{equation}
%  \sigma_A^{min}\otimes \sigma_B^{min}=\frac{I}{4}.\nonumber
%  \end{equation}
Therefore, the R\'{e}nyi total correlation of the Werner state for $\alpha \geq\frac{2}{3}~(\alpha\neq1)$ is given by
\begin{equation}
  \cal{I}^R_\alpha(\rho_{W})=2+ \frac{1}{\alpha-1}\log\frac{1}{4^{\alpha}}\big[(1+3p)^{\alpha}+3(1-p)^{\alpha}\big].
\end{equation}
Just like for the case of pure bipartite states, 
the R\'{e}nyi classical correlation is again independent of measurement basis, as
is expected from the property of rotational invariance of the Werner state.
% % We show this in the following 
% % proposition.
% % %\vskip10pt
% % 
% % \noindent {\bf Proposition:} \emph{For the Werner state, $\rho_W$, the total correlation of the post measurement state, $\rho'_W$, is independent of the measurement 
% % basis.}
% % %\vskip10pt
% % 
% % \noindent {\it Proof.}
% % Let
% % \begin{equation}
% %  \rho'_{W}=\sum_i(\mathbb{I}_A \otimes P_i) \rho_{W} (\mathbb{I}_A \otimes P_i),
% % \end{equation}
% % obtained by performing a measurement \(\{P_i\}\) on the Werner state. 
% % Suppose that we perform a measurement in another arbitrarily chosen basis, $P'_i$. 
% % The projectors $P'_i$ are necessarily connected to the $P_i$ through some unitary transformation, i.e,
% % $P'_i=UP_iU^\dagger$ \(\forall i\). Now the post-measurement state after the measurement \(\{P'_i\}\) is given by 
% % \begin{eqnarray}
% % \rho''_{W}&=&\sum_i(\mathbb{I}_A \otimes P'_i) \rho_{W} (\mathbb{I}_A \otimes P'_i) \nonumber\\
% % &=&\sum_iU\otimes U(\mathbb{I}_A \otimes P'_i)U^\dagger\otimes U^\dagger \rho_{W}U\nonumber\\
% % &\otimes& U(\mathbb{I}_A \otimes P'_i)U^\dagger\otimes U^\dagger\nonumber\\
% % &=&\sum_iU\otimes U(\mathbb{I}_A \otimes P'_i) \rho_{W}(\mathbb{I}_A \otimes P'_i)U^\dagger\otimes U^\dagger\nonumber\\
% % &=&U\otimes U\rho'_{W}U^\dagger\otimes U^\dagger.\nonumber
% % \end{eqnarray}
% % So, $\rho'_{W}$ and $\rho''_{W}$ are connected by a local unitary transformation. 
% % Therefore, their total correlations are equal. Hence, the result. \hfill\(\blacksquare\)
% % %\vskip10pt

Numerical observations also suggest that for $\alpha \geq\frac{1}{2}~(\alpha\neq1)$ and for any $p$, the
relative R\'{e}nyi entropy is minimum at $\sigma_A\otimes \sigma_B=\frac{I}{4}$ for
the post-measurement state corresponding to the Werner state. 
% \textcolor{red}{We have followed the same numerical techniques mentioned in Sec. 
% \ref{eijey!!}}. 
So the R\'{e}nyi classical correlation, in 
this range of $\alpha$, is given by
\begin{equation}
 \cal{J}^R_\alpha(\rho_{W})=2+ \frac{1}{\alpha-1}\log\frac{1}{4^{\alpha}}\big[2(1+p)^{\alpha}+2(1-p)^{\alpha}\big].
\end{equation}
Hence, the R\'{e}nyi quantum correlation of the Werner state for $\alpha \geq\frac{2}{3} ~(\alpha\neq1)$ is given by
\begin{equation}
 \cal{D}^R_\alpha(\rho_{W})= \frac{1}{\alpha-1}\log\left[\frac{(1+3p)^{\alpha}+3(1-p)^{\alpha}}{2(1+p)^{\alpha}
 +2(1-p)^{\alpha}}\right].
\end{equation}

\begin{figure}[htb]
\begin{center}
   \includegraphics[scale=0.5]{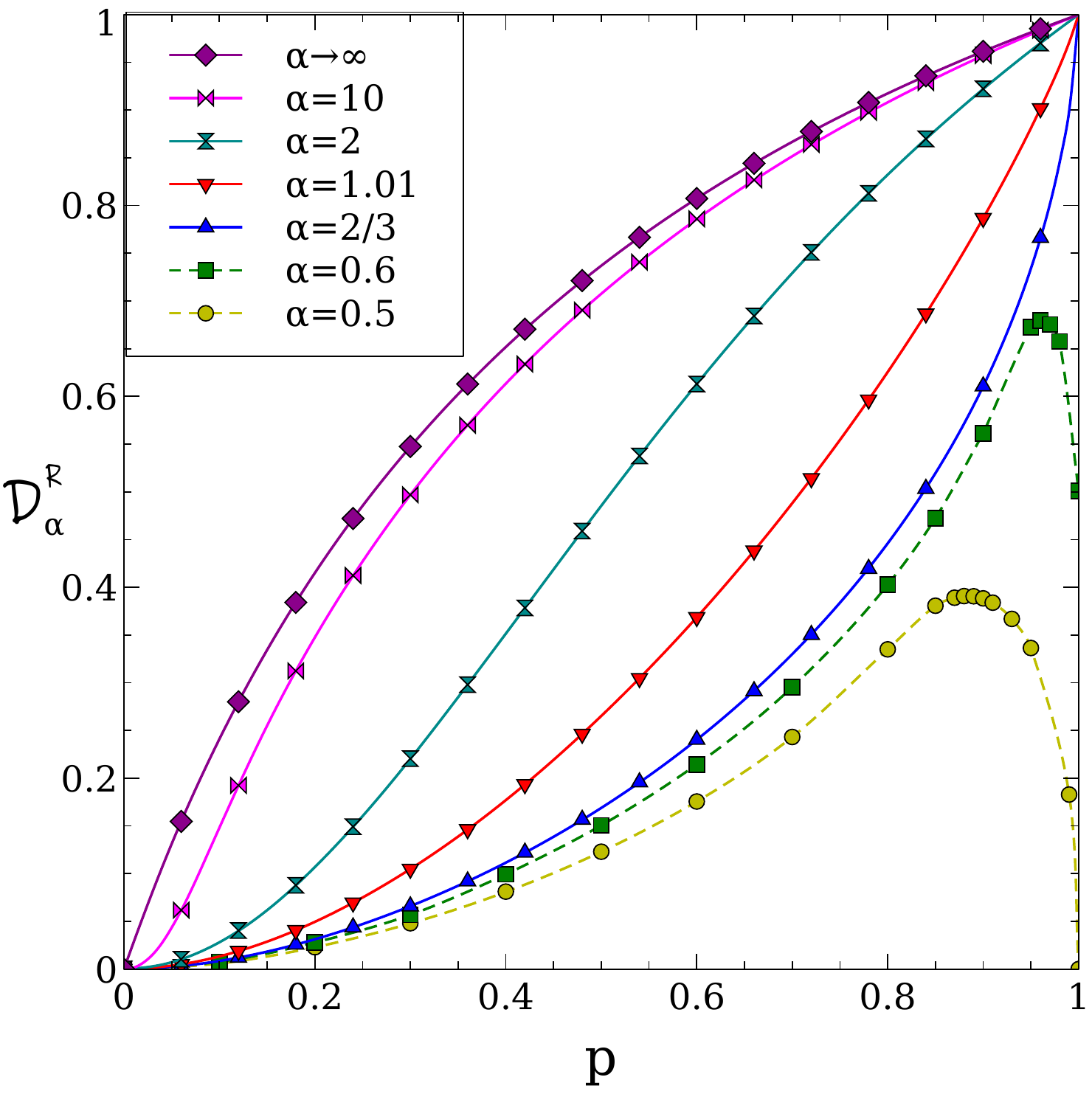}
\end{center}
\caption{(Color online.) R\'{e}nyi quantum correlation, $\cal{D}^R_\alpha$, with respect to $p$, of the Werner state, $\rho_{W}=
p|\psi^-\rangle\langle\psi^-| +(1-p)\frac{1}{4}I$,  for different $\alpha$. Both axes are dimensionless. }
\label{fig-werner}
\end{figure}

For $\frac{1}{2}\leq\alpha<\frac{2}{3}$, we find the R\'{e}nyi quantum correlation for the Werner states by numerical evaluation.
In Fig.~\ref{fig-werner}, we exhibit the R\'{e}nyi quantum correlation for the Werner states for different values of $\alpha$.
% It is to be noted that the singlet state no longer remains the state with maximal quantum discord for $\alpha<\frac{2}{3}$.

The R\'{e}nyi quantum correlation is maximum for the Werner state at $p=1$ for $\alpha \geq \frac{2}{3}$. 
% R\'{e}nyi quantum discord 
 The singlet state, and states that are local unitarily connected with it,
is therefore maximally R\'{e}nyi quantum correlated in that range of $\alpha$, among the Werner states. 
However, for $\frac{1}{2}\leq\alpha<\frac{2}{3}$, the Bell states are not the maximally 
R\'{e}nyi quantum correlated states. 
% In Fig.1. it is shown that in that range of $\alpha$ 
 In this range of $\alpha$, we get maximal quantum correlation among the Werner states, for a value of $p$ that is 
 different from unity.
 %$\neq 1$.
For example, for \(\alpha = 0.6\), we find that the state, \(\rho_W\), with mixing parameter \(p\approx0.96\) has the maximal quantum correlation among 
all Werner states. For \(\alpha =1/2\), the same is at \(p\approx0.88\).
 For $\alpha=\frac{1}{2}$, i.e., 
for min-entropy, the singlet has zero quantum correlation. Indeed, all pure states have vanishing min-quantum discord. We will visit this issue 
again in Sec. \ref{traditional}.

% We will discuss this in more detail in (\ref{minmax})
% 
% $\sigma_A\otimes \sigma_B$ is $\frac{I}{4}$ for total as well as classical correlation in case 
%  of R\'{e}nyi quantum discord for werner state for $\alpha \geq\frac{2}{3}, (\alpha\neq1)$. 
%  For any quantum state the optimum $\sigma_A\otimes \sigma_B$ is same for which the R\'{e}nyi and 
%  Tsallis relative entropy is minimum. So for werner state the we can calculate the total correlation
%  and classical correlation taking $\sigma_A^{min}\otimes \sigma_B^{min}=\frac{I}{4}$ for linear
%  entropy. So the total correlation for werner state for linear entropy is given by,
%  
% \begin{equation}
%   \cal{I}_L(\rho_{W})=\frac{1}{4}((1+3p)^{2}+3(1-p)^{2})-1,
% \end{equation}
% 
% and the classical correlation is given by,
% \begin{equation}
%   \cal{J}_L(\rho_{W})=\frac{1}{4}(2(1+p)^{2}+2(1-p)^{2})-1.
% \end{equation}
% Hence,
The linear quantum discord for the Werner state is 
\begin{equation}
  \cal{D}_L(\rho_{W})=\frac{1}{4}\left[(1+3p)^{2}+(1-p)^{2}-2(1+p)^{2}\right].
\end{equation}
% 
% Here we calculate 
The max-quantum discord can also be calculated similarly for the Werner state and is given by  
%  The total correlation is given by
%  \begin{equation}
%   \cal{I}_{max}(\rho_{W})=2+ \lim_{\alpha\rightarrow\infty}[\frac{1}{\alpha-1}\log_2(\frac{1}{4})^
%   {\alpha}((1+3p)^{\alpha}+3(1-p)^{\alpha})].\nonumber
% \end{equation}
% Calculating the limit we get
%  \begin{equation}
%   \cal{I}_{max}(\rho_{W})=2+ \log_2\frac{(1+3p)}{4}.
% \end{equation}
% Similarly classical correlation is given by
% \begin{equation}
%   \cal{J}_{max}(\rho_{W})=2+ \lim_{\alpha\rightarrow\infty}[\frac{1}{\alpha-1}\log_2(\frac{1}{4})^
%   {\alpha}(2(1+p)^{\alpha}+2(1-p)^{\alpha})].\nonumber
% \end{equation}
% Calculating the limit we get
% \begin{equation}
%   \cal{J}_{max}(\rho_{W})=2+ \log_2\frac{(1+p)}{4}.
% \end{equation}
% 
% Hence,
\begin{equation}
  \cal{D}_{max}(\rho_{W})= \log\left[\frac{(1+3p)}{(1+p)}\right].
\end{equation}
 We have numerically evaluated the min-quantum discord for the Werner state (see Fig.~\ref{fig-werner}).  
 %\vskip 10pt
 
\begin{figure}[htb]
\begin{center}
   \includegraphics[scale=0.5]{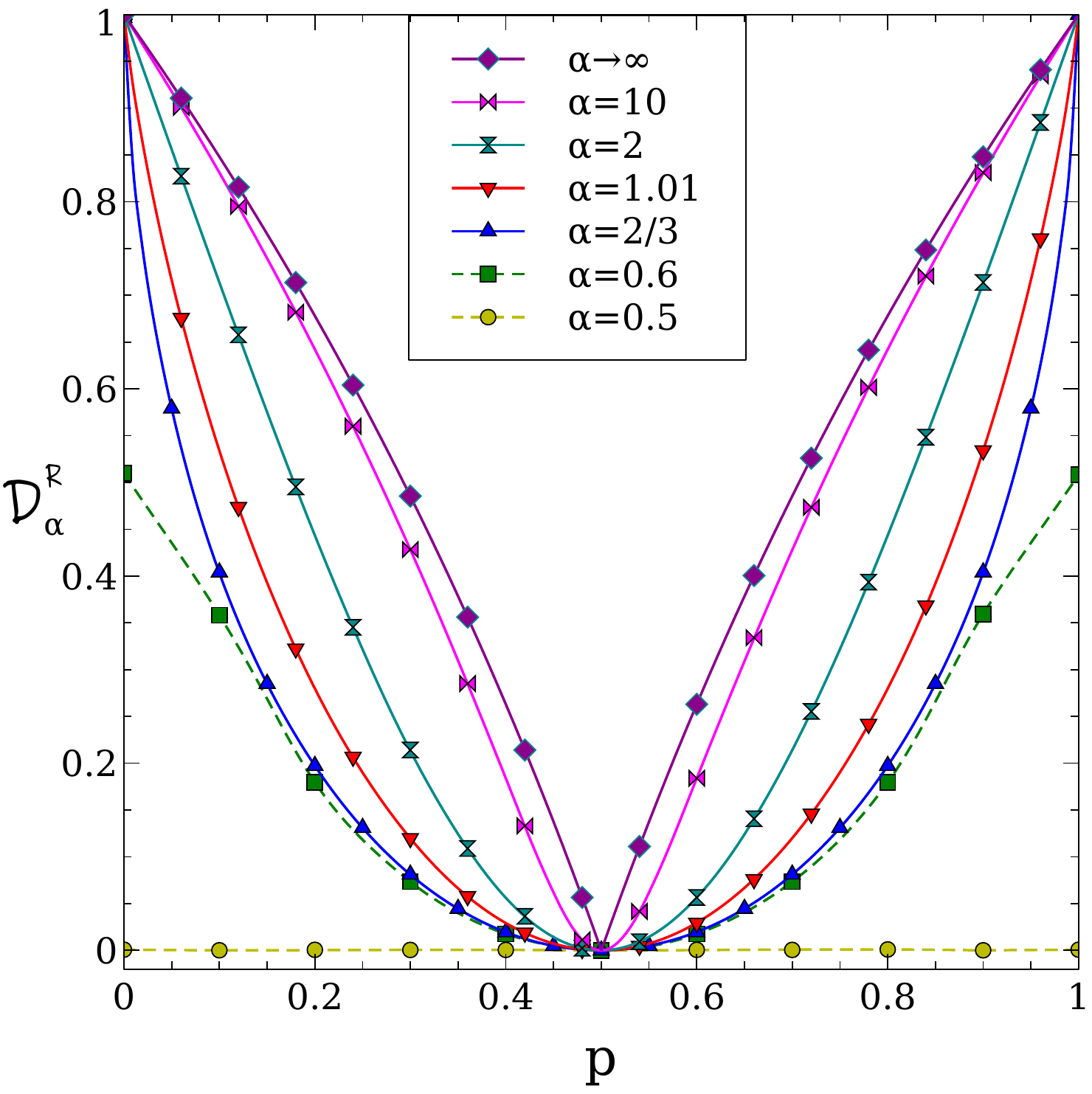}
\end{center}
\caption{(Color online.) R\'{e}nyi quantum correlation, $\cal{D}^R_\alpha$, with respect to $p$, of the Bell mixture, 
$\rho_{BM}=p|\phi^+\rangle\langle\phi^+| +(1-p)|\phi^-\rangle\langle\phi^-|$,   for different values of $\alpha$. 
Both axes are dimensionless.}
\label{fig-bellmix}
\end{figure}

\noindent{\bf (ii) Bell Mixture:} We consider a mixture of two Bell states, given by 
\begin{equation}
 \rho_{BM}=p|\phi^+\rangle\langle\phi^+| +(1-p)|\phi^-\rangle\langle\phi^-|, \nonumber
\end{equation}
where $|\phi^+\rangle=\frac{1}{\sqrt{2}}(|00\rangle+|11\rangle)$, $|\phi^-\rangle=\frac{1}{\sqrt{2}}
(|00\rangle-|11\rangle)$ and \(0\leq p \leq 1\).
Numerical observations suggests that 
\begin{equation}
 \cal{I}^{\varGamma}_\alpha(\rho_{BM})=\tilde{S}_\alpha^{\varGamma}\left(\rho_{BM}\vert\vert \frac{I}{4}\right),\nonumber
\end{equation}
for $\alpha \geq\frac{2}{3}~(\alpha\neq1)$.
Hence, in this range of $\alpha$,
\begin{equation}
 \cal{I}^{R}_\alpha(\rho_{BM})= 2+ \frac{1}{\alpha-1}\log\big[p^\alpha+(1-p)^\alpha\big].
\end{equation}
We have found numerically that if one performs measurement in the $\{\vert0\rangle$, $\vert1\rangle\}$ basis,
% and then 
the relative entropy of the post-measurement state with $\frac{I}{4}$ gives the R\'{e}nyi classical
correlation for the entire range of $\alpha$, i.e., for $\alpha\in (\frac{1}{2},1) \cup (1, \infty)$, and it is equal to unity for any $p$ and $\alpha$.
Hence for $\alpha \geq\frac{2}{3}~(\alpha\neq1)$,
\begin{equation}
 \cal{D}^{R}_\alpha(\rho_{BM})= 1+ \frac{1}{\alpha-1}\log\left[p^\alpha+(1-p)^\alpha\right].
\end{equation}
The linear quantum discord for this state is given by
% \begin{equation}
%  4(p^2+(1-p)^2)-1\nonumber \\
%  =8(p^2-p)+3
% \end{equation}
% and classical correlation is one.
% Hence,
\begin{equation}
\cal{D}_L(\rho_{BM})=8(p^2-p)+2.
\end{equation}
Similarly,
\begin{eqnarray}
\cal{D}_{max}(\rho_{BM})
% = 1+ \lim_{\alpha\rightarrow\infty}\frac{1}{\alpha-1}\log_2[p^\alpha+(1-p)^\alpha] \nonumber \\ 
 =1+\log\left[\mbox{max}\{p,1-p\}\right].
\end{eqnarray}
%\textcolor{red}{The numerical techniques used here are similar to that mentioned in Sec. \ref{eijey!!}}.
%For $\frac{1}{2}\leq\alpha<\frac{2}{3}$, we have numerically evaluated the R\'{e}nyi quantum correlation for $\rho_{BM}$.
In Fig.~\ref{fig-bellmix}, the R\'{e}nyi quantum correlations for $\rho_{BM}$ is depicted for different values of $\alpha$.
%\vskip 10pt

\begin{figure}[htb]
\begin{center}
   \includegraphics[scale=0.5]{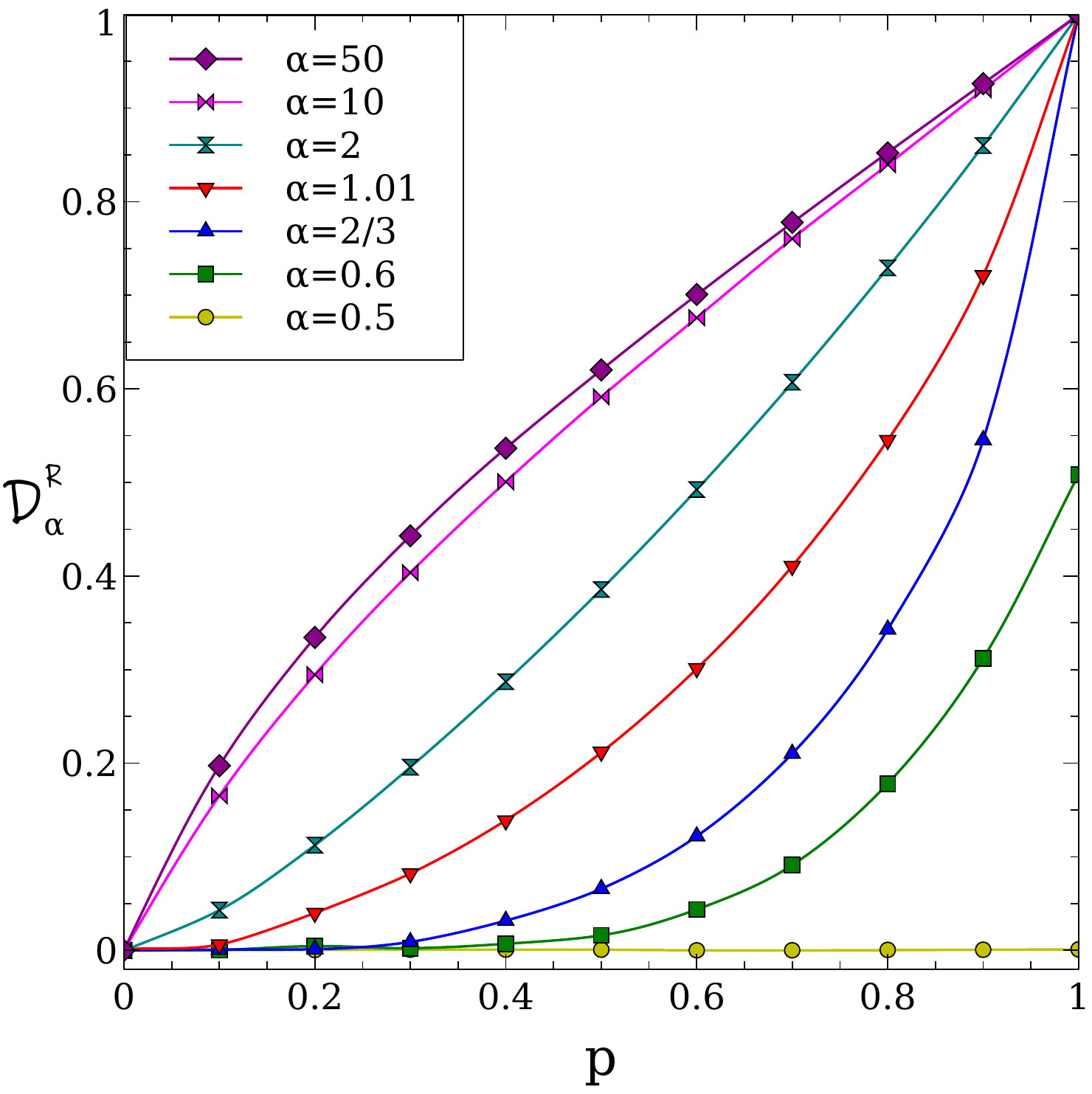}
\end{center}
\caption{(Color online.) R\'{e}nyi quantum correlation, $\cal{D}^R_\alpha$, with respect to $p$, of 
$\rho_{BN}=p|\phi^+\rangle\langle\phi^+| +(1-p)|00\rangle\langle00|$,   for different $\alpha$. 
Both axes are dimensionless.}
\label{fig-bellnoise}
\end{figure}

\noindent{\bf(iii) Mixture of Bell state and a Product State:} Consider the state given by
\begin{equation}
 \rho_{BN}=p|\phi^+\rangle\langle\phi^+| +(1-p)|00\rangle\langle00|. \nonumber
\end{equation}
The R\'{e}nyi quantum correlation is calculated numerically, and
in Fig.~\ref{fig-bellnoise}, we plot it for $\rho_{BN}$, for different values of $\alpha$.
%\textcolor{red}{We have performed the numerical studies by the same techniques mentioned in Sec. \ref{eijey!!}}.
\subsection{Sandwiched vs Traditional Relative Entropies}
\label{traditional}
Until now, in this section, we have used the sandwiched relative entropy distances to define the R\'{e}nyi and 
Tsallis quantum correlations. We now briefly consider the traditional
variety for defining quantum correlation, and discuss some of its implications.
In the preceding subsections, 
we have observed anomalous behavior of the R\'{e}nyi quantum correlation in the range $\frac{1}{2}\leq\alpha<\frac{2}{3}$ for pure states,
as well as in certain families of mixed states in the neighborhood of pure states. 
%The anomalies are 
%due to the sandwiched form of the relative entropy. 
In these cases, we have, e.g., seen that the Bell states are not the maximally R\'{e}nyi 
quantum correlated state for $\alpha<\frac{2}{3}$ and
at $\alpha=\frac{1}{2}$, i.e, for the min- entropy, all pure states have vanishing quantum correlations.

We can also define quantum correlations with the traditional 
relative R\'{e}nyi and Tsallis
entropies. The properties (1-4) discussed in Sec. \ref{gre}, are also followed by both the traditional relative entropies~\cite{furuchi}, 
but the data processing inequality holds for $\alpha\in [0,1) \cup (1, 2]$~\cite{traddpi}. We can therefore define quantum 
correlation with traditional relative entropy distances for this range of $\alpha$.
If we consider 
the traditional relative entropies, then we do not see any anomalous behavior of the R\'{e}nyi quantum correlation.
But from the traditional version of the relative R\'{e}nyi entropy, we do not get the min- entropy. Moreover, in \cite{ogawa}, the authors have argued 
that the sandwiched relative R\'{e}nyi entropy is operationally relevant in the strong converse problem of quantum hypothesis testing for $\alpha>1$, 
but for $\alpha<1$, the traditional version is more relevant from an operational point of view. The anomalous behavior of the quantum correlation with 
the sandwiched relative entropy distances seems to indicate that to define quantum correlation for $\alpha<1$, the more 
appropriate candidates are the traditional relative entropies.
% may be
% more appropriate choices.
Here we discuss about the traditional R\'{e}nyi
quantum correlation for two-qubit pure states and the Werner state.\\
%\vskip10pt
\noindent{\bf(i) Pure States:}
Numerical observations similar to the case with the sandwiched variety, give us that the total correlation of a two-qubit pure state, 
$|\psi_{AB}\rangle=\displaystyle\sum_{i=0}^1 \sqrt{\lambda_i} |i_{A}i_{B}\rangle$, for traditional relative R\'{e}nyi entropy, 
with $\alpha\in(\frac{1}{2},1)$, is given by
\begin{eqnarray}
\cal{I}^{TR}_\alpha(\vert\psi_{AB}\rangle)&=&\min_{\{a\}}\frac{1}{\alpha-1} \log\Big[\lambda a^{2(1-\alpha)} \nonumber \\
&+& (1-\lambda)(1-a)^{2(1-\alpha)}\Big], 
\end{eqnarray}
where $0\leq a\leq1$, and the value of $a$ is obtained from the condition
%  Now we have to find the value of $p$ in terms of $\alpha$ and $a$ from the minimization condition. 
%  From the minimization condition, for $\alpha\geq1$ we have
 \begin{equation}
  \frac{1}{a}=\left(\frac{\lambda}{1-\lambda}\right)^{\frac{1}{1-2\alpha}} +1.
 \end{equation}
% %  For $\alpha\in(0,\frac{1}{2}]$, the same is given by
% % \begin{equation}
% %  \cal{I}^{TR}_\alpha(\vert\psi_{AB}\rangle)= -\frac{1}{\alpha-1}\log\big[\max\{\lambda,(1-\lambda)\}\big].
% % \end{equation}
  \begin{figure}[htb]
\begin{center}
   \includegraphics[scale=0.5]{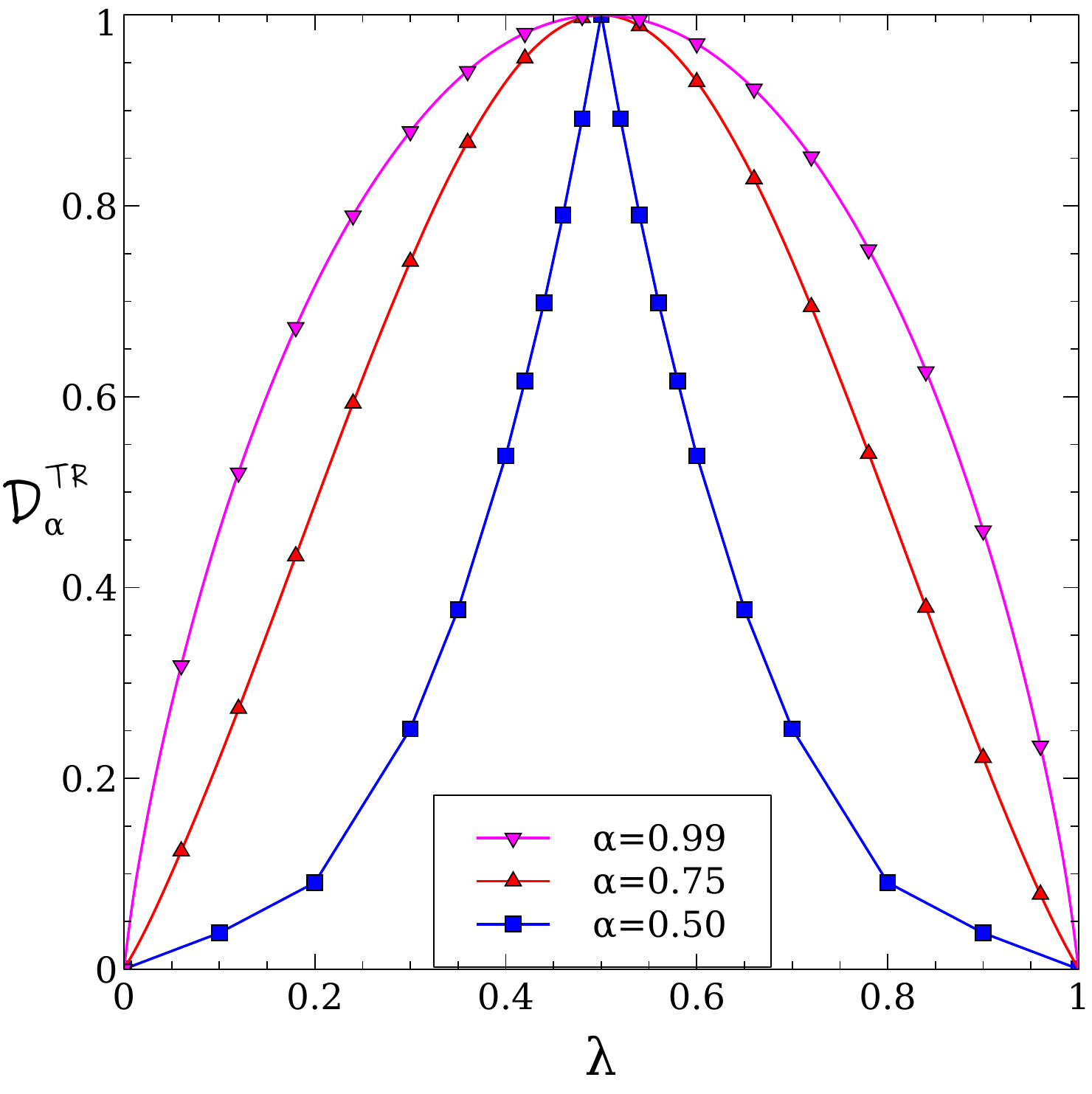}
\end{center}
\caption{(Color online.) Traditional R\'{e}nyi quantum correlation, $\cal{D}^{TR}_\alpha$, with respect to $\lambda$, of $\vert\psi_{AB}\rangle=
\sqrt{\lambda}|00\rangle+\sqrt{(1-\lambda)}|11\rangle$  for different $\alpha$. Both axes are dimensionless. }
 \label{fig-tradpure}
 \end{figure}
 The classical correlation in the traditional case in computed numerically. The numerical computation is performed by 
 the same numerical recipe as mentioned in Sec. \ref{eijey!!}.
 
%For the same range of $\alpha$, the classical correlation for the traditional one is same as the sandwiched relative R\'{e}nyi entropy. For 
%$\alpha\in[0,\frac{1}{2}]$, the traditional R\'{e}nyi quantum correlation is independent of $\alpha$ and is given by
%\begin{equation}
% \cal{D}^{TR}_\alpha(\vert\psi_{AB}\rangle)= -\log\big[\max\{\lambda,(1-\lambda)\}\big].
%\end{equation}
In Fig.~\ref{fig-tradpure}, we have plotted the $\cal{D}^{TR}_\alpha(\vert\psi_{AB}\rangle)$, for different values of $\alpha$. No anomalous 
behavior can be seen, and the maximally entangled states have maximal quantum correlations.\\
%\vskip10pt
\noindent{\bf(ii) Werner States:}
 Like in the sandwiched version, exploiting the rotational invariance and symmetry of the Werner state, it can be shown analytically that the total correlation
 of the Werner state for the traditional relative R\'{e}nyi entropy, for $\alpha\in[\frac{1}{2},1)$, is given by
 \begin{equation}
  \cal{I}^{TR}_\alpha(\rho_{W})=2+ \frac{1}{\alpha-1}\log\frac{1}{4^{\alpha}}\big[(1+3p)^{\alpha}+3(1-p)^{\alpha}\big].
\end{equation}
The classical correlation of the Werner state is also measurement basis independent for the traditional version, like the sandwiched one.
% , and one can prove 
% it like the way we proved it for the sandwiched one.
% From numerical studies, w
We get that the classical correlation, in this range,  is given by 
\begin{equation}
 \cal{J}^{TR}_\alpha(\rho_{W})=2+ \frac{1}{\alpha-1}\log\frac{1}{4^{\alpha}}\big[2(1+p)^{\alpha}+2(1-p)^{\alpha}\big].
\end{equation}
\begin{figure}[htb]
\begin{center}
   \includegraphics[scale=0.5]{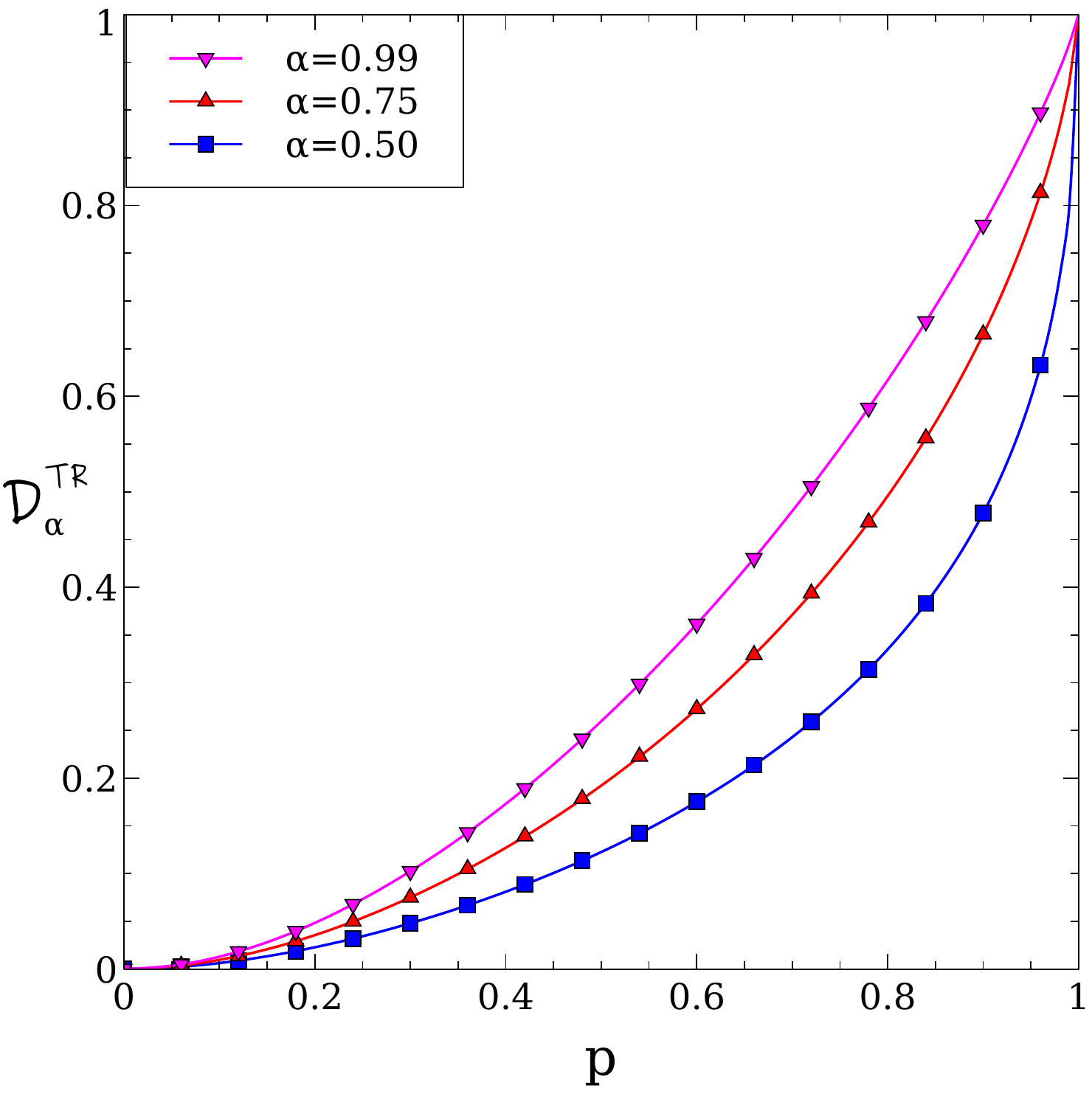}
\end{center}
\caption{(Color online.) Traditional R\'{e}nyi quantum correlation, $\cal{D}^{TR}_\alpha$, with respect to $p$ of the Werner state, $\rho_{W}=
p|\psi^-\rangle\langle\psi^-| +(1-p)\frac{1}{4}I$,  for different $\alpha$. Both axes are dimensionless. }
\label{fig-tradwerner}
\end{figure}
%\textcolor{red}{The same numerical techniques, mentioned in Sec. \ref{eijey!!}, are followed here also}.
The forms of the total and classical correlations, in this case, are equivalent to those in the sandwiched version. But
here, the range of $\alpha$ is different. Hence, for $\alpha\in[\frac{1}{2},1)$, the traditional R\'{e}nyi quantum correlation for the Werner state is given by
\begin{equation}
 \cal{D}^{TR}_\alpha(\rho_{W})= \frac{1}{\alpha-1}\log\left[\frac{(1+3p)^{\alpha}+3(1-p)^{\alpha}}{2(1+p)^{\alpha}
 +2(1-p)^{\alpha}}\right].
\end{equation}

% For $\alpha<\frac{1}{2}$, we have numerically evaluated the traditional R\'{e}nyi quantum correlation for the 
% Werner state and no anomalous behavior is noticed for  $\alpha\in(0,1)$. 
In Fig.~\ref{fig-tradwerner}, we have plotted  
the $\cal{D}^{TR}_\alpha(\rho_{W})$, for different values of $\alpha$.
%\vskip20pt

\section{Application: Detecting criticality in quantum ising model}
\label{qpt}
In this section, we show that the R\'{e}nyi and Tsallis quantum correlations can be applied to detect cooperative phenomena in quantum many-body systems. 
% We will find that 
% the scaling exponents of generalized quantum discord near critical points are different 
% from those of quantum correlation measures. 
%on some physical systems, specifically on spin models. 
% More specifically, we study the scaling behavior of the generalized quantum discord in 
% the one-dimensional quantum Ising model~\cite{Barouch-McCoy}. 
Let us consider a  system of  $N$ quantum spin-1/2 particles, described by the one-dimensional
quantum Ising model~\cite{Barouch-McCoy}. 
Such models can be simulated by using  ultracold gases  in a controlled way in the 
laboratories~\cite{Strotzky, maciej}, and is also known to describe 
Hamiltonians of materials~\cite{Mschechter}.
The Hamiltonian for this system is given by
\begin{equation}
\label{eq_XY_H}
 H = J \sum_{i=1}^{N} \sigma^x_i \sigma^x_{i+1}  + h \sum_{i=1}^{N} \sigma_i^z,
\end{equation}
where $J$ is the coupling constant for the nearest neighbor interaction, \(\sigma\)'s are the Pauli spin matrices, 
and \(h\) represents the external transverse magnetic field applied across the system. 
Periodic boundary condition is assumed.
% $\gamma=1$ corresponds to the transverse Ising model.
%Moreover, we impose the periodic boundary condition. 
The Hamiltonian can be diagonalized by applying Jordan-Wigner, Fourier, and Bogoliubov 
transformations~\cite{Barouch-McCoy}. At zero temperature, it undergoes 
a quantum phase transition (QPT) driven by the transverse magnetic field at 
$\lambda\equiv\dfrac{h}{J}=\lambda_c\equiv1$~\cite{Barouch-McCoy}. Such a transition has been detected by using 
different order parameters~\cite{Barouch-McCoy,Mefisher}, including quantum correlation measures like 
concurrence~\cite{Aosterloh}, 
geometric measures~\cite{Ashimony,Asde,Abiswas}, and quantum discord~\cite{Rdillenschneider}.

\begin{figure}[]
\begin{center}
   \includegraphics[scale=0.33]{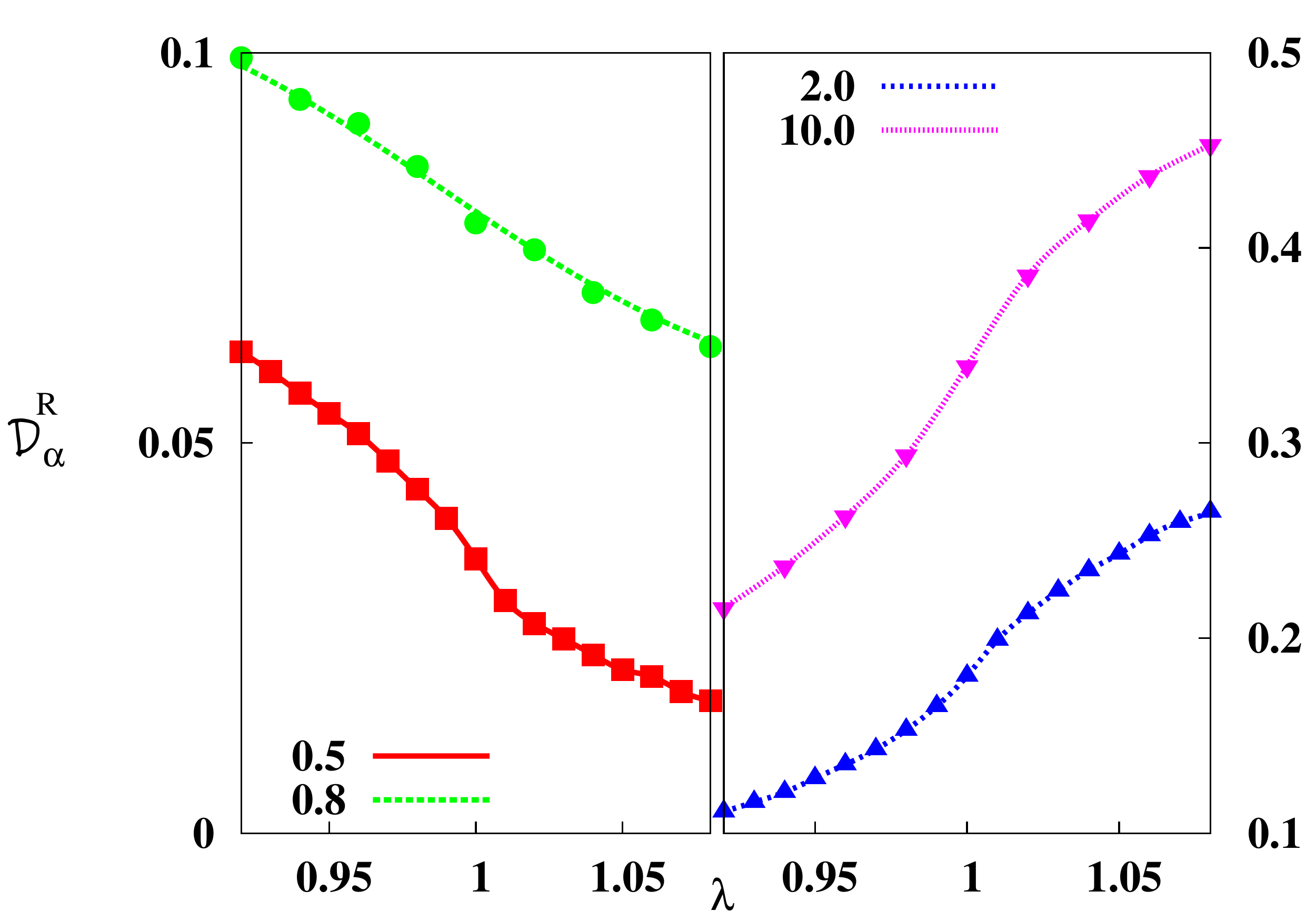}
\end{center}
\caption{(Color online.) Detecting quantum phase transitions with R\'{e}nyi quantum correlations. R\'{e}nyi quantum correlation,
$\cal{D}_\alpha^R$, with respect to $\lambda$, of the nearest neighbor bipartite density matrix at zero temperature,
%$\rho_{AB}$, 
for different values of $\alpha$. The legends indicate the values of $\alpha$.
Both axes are dimensionless.}
% The symbols indicate the points where the evaluations have been carried out explicitly. The curves are then drawn as 
% a guide to the eye.

\label{fig-ising-inf}
\end{figure}

\begin{figure}[]
\begin{center}
   \includegraphics[scale=0.35]{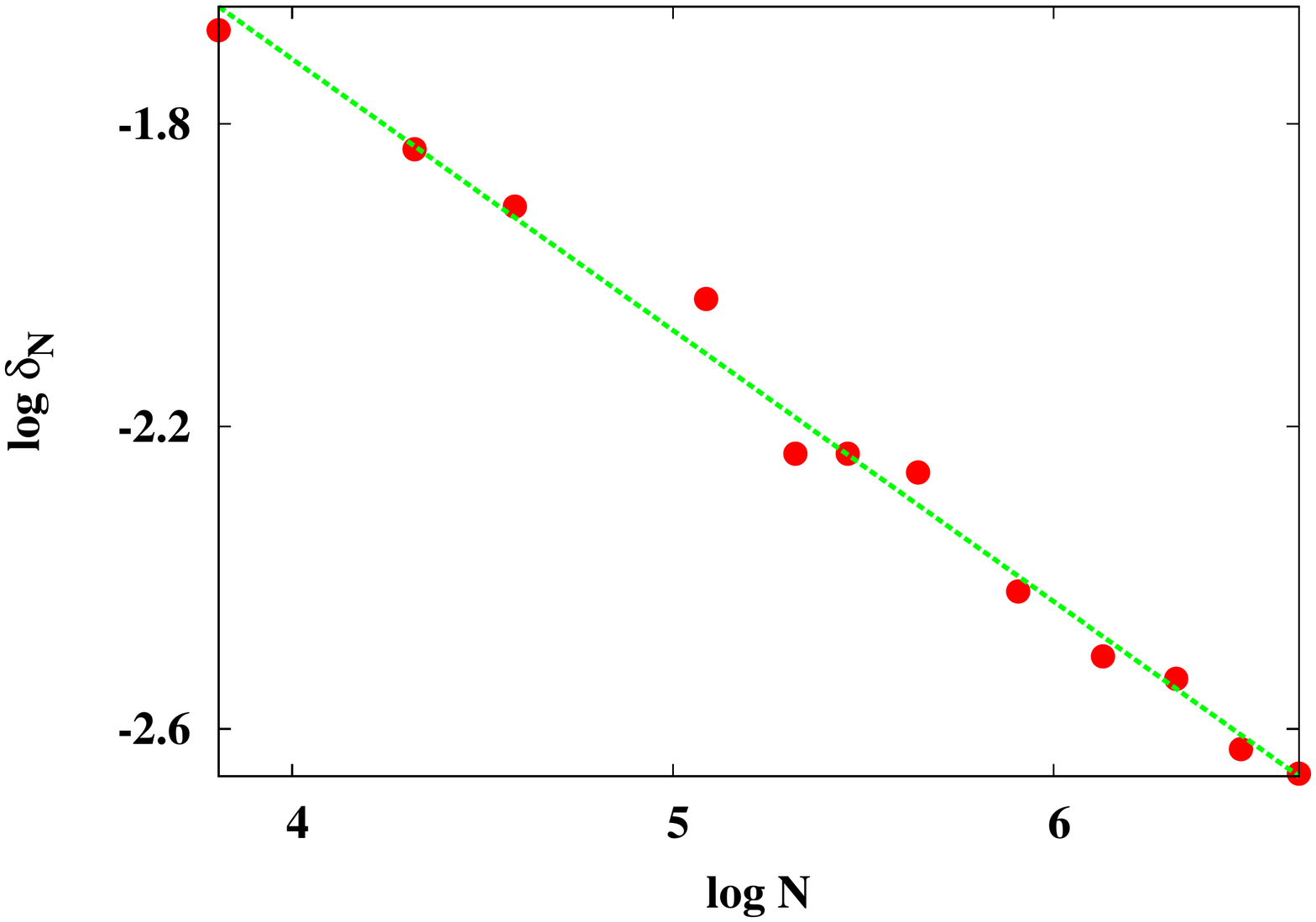}
\end{center}
\caption{(Color online.)
% Divergence of $\frac{d\cal{D}_2^R}{d\lambda}$ at the QPT point for the infinite transverse quantum Ising model.
% %Detecting quantum phase transitions with R\'{e}nyi quantum correlations. R\'{e}nyi quantum correlation,
% %$\cal{D}_\alpha^R$, with respect to $\lambda$, of the nearest neighbor bipartite density matrix at zero temperature,
% %$\rho_{AB}$, 
% %for different values of $\alpha$. The legends indicate the values of $\alpha$.
% Both axes are dimensionless. Inset: 
Scaling analysis of full-width at half maxima, $\delta_N$, for \(\cal{D}_2^R\).
Both axes are dimensionless.}
%\sout{The vertical axis is dimensionless, while the horizontal one is in log of the number of particles.}}
% The symbols indicate the points where the evaluations have been carried out explicitly. The curves are then drawn as 
% a guide to the eye.
\label{fig-ising-divergence}
\end{figure}

We now investigate the behavior of the R\'{e}nyi and Tsallis quantum correlations of the nearest neighbor density matrix (reduced density matrix of two neighboring spins)
at zero temperature,
near the quantum critical point. Note that we have reverted back to the sandwiched version of the relative 
entropies in this section.
The nearest neighbor bipartite density matrix, $\rho_{AB}$, of the ground state of the Hamiltonian given by Eq.~(\ref{eq_XY_H}),
represented by $\rho_{AB}$, 
can be written~\cite{Barouch-McCoy} in terms of the diagonal two-site correlators and the average magnetization
in $z$-direction. The density matrix, $\rho_{AB}$, in the 
thermodynamic limit of \(N\to \infty\), is given by\\\\
\hspace*{1cm}$\small{\rho_{AB}=\begin{pmatrix} 
  \alpha_{+}+\dfrac{M_z}{2} & 0 & 0 & \beta_{-}\\ 
  0 & \alpha_{-} & \beta_{+} & 0\\
  0 & \beta_{+} &  \alpha_{-} & 0\\
  \beta_{-} & 0 & 0 & \alpha_{+}-\dfrac{M_z}{2}
\end{pmatrix}}$\\\\
where $\alpha_{\pm}=\dfrac{1}{4}(1\pm T_{zz})$, $\beta_{\pm}=\dfrac{T_{xx}\pm T_{yy}}{4}$ 
with $T_{ij}=\mbox{tr}(\sigma_i\otimes\sigma_j\rho_{AB})$
and $M_z=\mbox{tr}(\mathbb{I_A}\otimes\sigma_z \rho_{AB})$.
The correlations and transverse magnetization, for the zero-temperature state, 
are given by~\cite{Barouch-McCoy} 
\begin{eqnarray}
T^{xx}(\lambda)&=&G(-1, \lambda),\nonumber\\
T^{yy}(\lambda)&=&G(1, \lambda),\\
T^{zz}(\lambda)&=& [M^z(\lambda)]^2-G(1, \lambda)G(-1, \lambda),\nonumber
\label{eq:Tijeq}
\end{eqnarray}
where 
% $G(R, \lambda)$ are 
\begin{eqnarray}
G(R, \lambda) =  \frac{1}{\pi}\int^\pi_0d\phi \frac{( \sin(\phi R)\sin \phi - \cos\phi (\cos\phi -\lambda))} 
{\Lambda(\lambda)} \nonumber\\
% &\times &( \sin(\phi R)\sin \phi - \cos\phi (\cos\phi -\lambda)) \nonumber\\
\end{eqnarray}
and 
\begin{eqnarray}
M^z(\lambda) &=& -\frac{1}{\pi} \int_0^\pi d\phi \frac{ (\cos\phi -\lambda)}{\Lambda(\lambda)}. \nonumber \\
 \label{eq:Mzeq}
\end{eqnarray}
Here 
\begin{equation}
\Lambda(x)= \left\{\sin^2\phi~+~[x-\cos\phi]^2\right\}^{\frac{1}{2}},
\end{equation}
and 
\begin{equation}
 \lambda = \frac{h}{J}. 
%\quad \tilde{\beta} = \beta J.
\end{equation}
Note that \(\lambda\)
%and \(\tilde{\beta}\) are  
is a 
dimensionless variable. The R\'{e}nyi and Tsallis quantum correlations are calculated for the state, $\rho_{AB}$,
for different values of $\alpha$. In Fig.~\ref{fig-ising-inf}, we plot the R\'{e}nyi quantum correlation
 as a function of $\lambda$ for different values of $\alpha$. 
 QPT corresponds to a point of inflexion in the \(\cal{D}^\varGamma_\alpha\) versus \(\lambda\) curve and $\frac{d\cal{D}_\alpha^\varGamma}{d\lambda}$ 
 %also 
 diverges there.
 We claim that the derivatives of the R\'{e}nyi (and the Tsallis) discords do diverge at the critical point. The seeming finiteness of the 
 derivative at the critical point has to do 
with the finite spacing of the variable $\lambda$. To see this, we perform a finite-size scaling analysis
of the full width at half maxima, of the peak tat is obtained around the critical point for finite size (see Fig.~\ref{fig-ising-divergence}).

 %at the QPT point 
 %(see Fig.~\ref{fig-ising-divergence}). 
% It can be clearly seen from the 
%  figure that $\frac{d\cal{D}_\alpha^R}{d\lambda}$, has a minima at the 
%  QPT for $\alpha<1$,
%  while it has a maxima at the QPT for $\alpha>1$. The same feature persists for the Tsallis 
%  quantum correlation.
 This feature is distinctly different from the variation of the derivative of the quantum discord with respect to $\lambda$ around the QPT point, which 
  exhibits a point of inflexion at $\lambda=1$~\cite{Rdillenschneider} (cf. \cite{Lidar-qpt}). 
  It is only the second derivative of quantum discord with respect to $\lambda$,
  which diverges at the QPT point.
  %, signaling a phase transition. 
  This is an uncomfortable and intriguing 
  feature of quantum discord, and is not shared by e.g. 
  the concurrence at the same quantum critical point \cite{Aosterloh}.
  Therefore it is advantageous to use the 
  R\'{e}nyi and Tsallis quantum correlations to detect phase transitions and other collective phenomena in 
  quantum many body systems, in comparison to quantum discord.
% $F_G$ corresponds to the maximum eigenvalue of the density matrix $\rho_{AB}$. 
% $F_L$ is obtained by numerical maximization of the density matrix $\rho_{AB}$ with respect to the product states in $\mathbb{C}^2 \otimes \mathbb{C}^2$.
%  We plot the derivative of $S_P$ with respect to $\lambda$, \emph{i.e.} $\frac{dS_P}{d\lambda}$, against $\lambda$, in Fig.~\ref{fig-aniso},
% for different values of the anisotropy parameter $\gamma$.
% The divergence of the derivative at $\lambda=\lambda_c \equiv 1$ clearly signals the quantum phase transition in this model, indicating that 
% shared purity can be used as a physical quantity to detect 
% quantum phase transitions. \\

% Therefore, it can be seen that the shared purity of the two-party density matrix of an 
% $N$-party state is sufficient to detect the quantum phase transition in this model.
\noindent \textbf{Finite-size scaling~:} The R\'{e}nyi and Tsallis  quantum correlations are shown in Fig. \ref{fig-ising-inf} to detect phase 
transitions in infinite systems. Ultracold gas realization of such phenomena, however, can simulate the corresponding Hamiltonian
for a finite number of spins~\cite{Rislam}.
The quantum Ising model, which has been briefly described earlier in this section, can 
also be solved for finite-size systems~\cite{Barouch-McCoy}.
We calculate the quantum correlations of nearest neighbor spins for finite spin 
chains using both
the Tsallis and R\'{e}nyi entropies. We find that the quantum correlations
detect the transition in finite-size systems too. Again, the transition point corresponds to points of inflexion in the \(\cal{D}^\varGamma_\alpha\) versus \(\lambda\) curves,
and narrow bell-shaped peaks in the $\frac{d\cal{D}_\alpha^\varGamma}{d\lambda}$ versus \(\lambda\) curves, for different values of \(N\).
%shows a narrow bell-shaped peak 
%at the QPT point. 
The bell-shaped curves become more narrow and peaked with the increase of number of spins. We perform a finite-size 
scaling analysis of full-width at half maxima, $\delta_N$, of the $\frac{d\cal{D}_\alpha^\varGamma}{d\lambda}$ versus \(\lambda\) curves, 
and the scaling exponent is e.g. -0.36 for \(\cal{D}_2^R\) (see Fig.~\ref{fig-ising-divergence}).
The exponent is a measure of the rapidity with which the narrow bell-shaped peak tends to show a divergence
%QPT point, $\lambda_c^N,$ 
with the increment in system size $N$. The $\log-\log$ scaling between the size, $N$, and the width, $\delta_N$, clearly indicates divergence
of the derivative at
infinite $N$.
%approaches the QPT point, $\lambda_c$, of the infinite system, 
%as a function of $N$.Please the  Fig. 9 of the current manuscript. We have now mentioned this more carefully in the revised manuscript. size 

We also perform finite-size scaling analyses of the \(\lambda_c^N\), the value of \(\lambda\) for which the derivatives of the 
R\'{e}nyi (or Tsallis) quantum correlations with respect to \(\lambda\) has a maximum for a system of \(N\) spins,
for several different values of $\alpha$, and obtain the corresponding scaling exponents. 
\begin{figure}[h]
\begin{center}
   \includegraphics[scale=0.35]{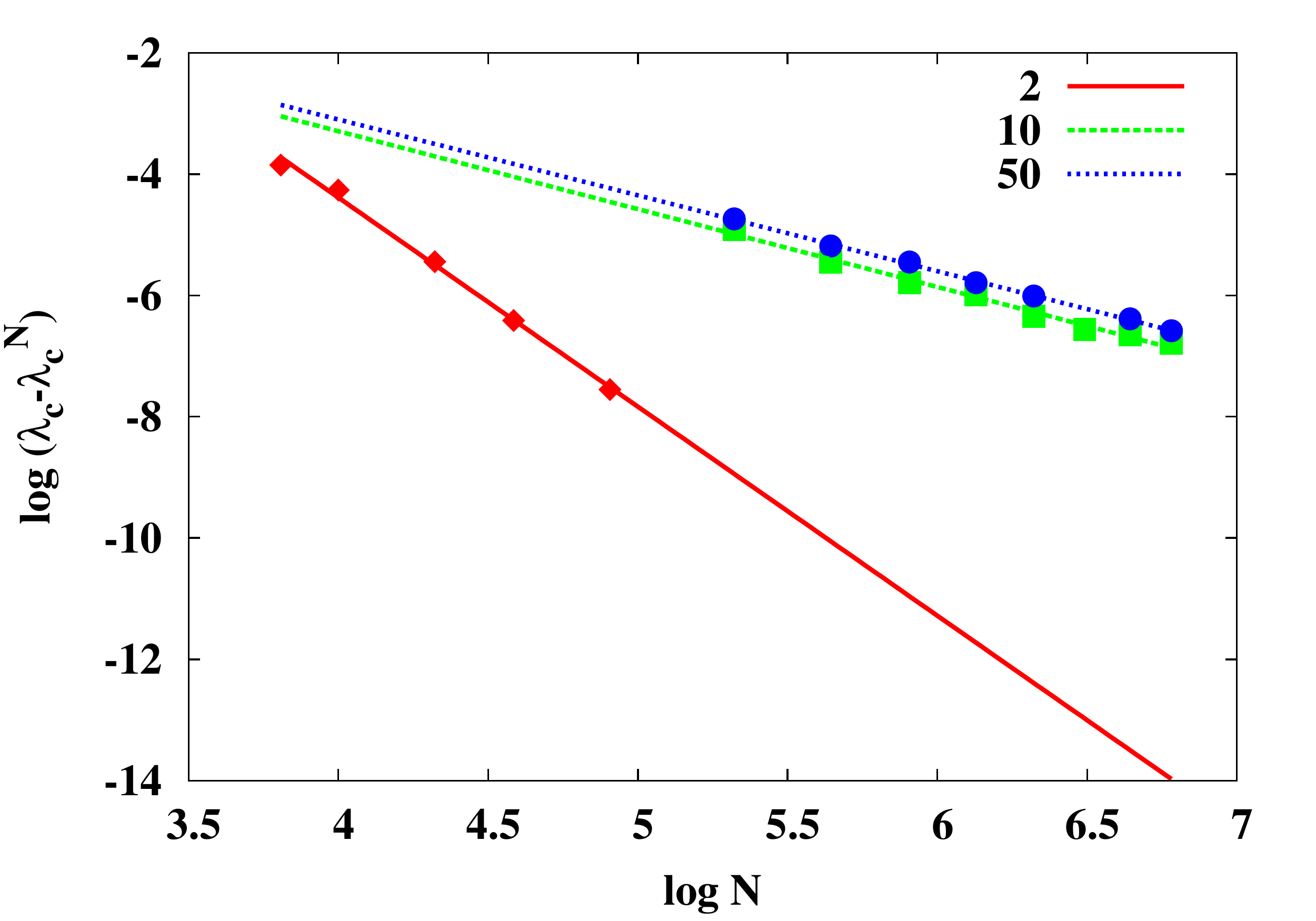}
\end{center}
\caption{(Color online.) Scaling analysis of R\'{e}nyi quantum correlation, $\cal{D}_\alpha^R$, for
different values of $\alpha$, in the one-dimensional quantum Ising model. The legends indicate the values of $\alpha$.
Both axes are dimensionless.}
%\sout{The vertical axis is dimensionless, while the horizontal one is in log of the number of spins.}}
\label{fig-scaling-ren}
\end{figure}
 The exponent is a measure of the rapidity with which the QPT point, $\lambda_c^N,$ in a finite
size system of size $N$, approaches the QPT point, $\lambda_c$, of the infinite system, as a function of $N$.

\begin{figure}[h]
\begin{center}
   \includegraphics[scale=0.35]{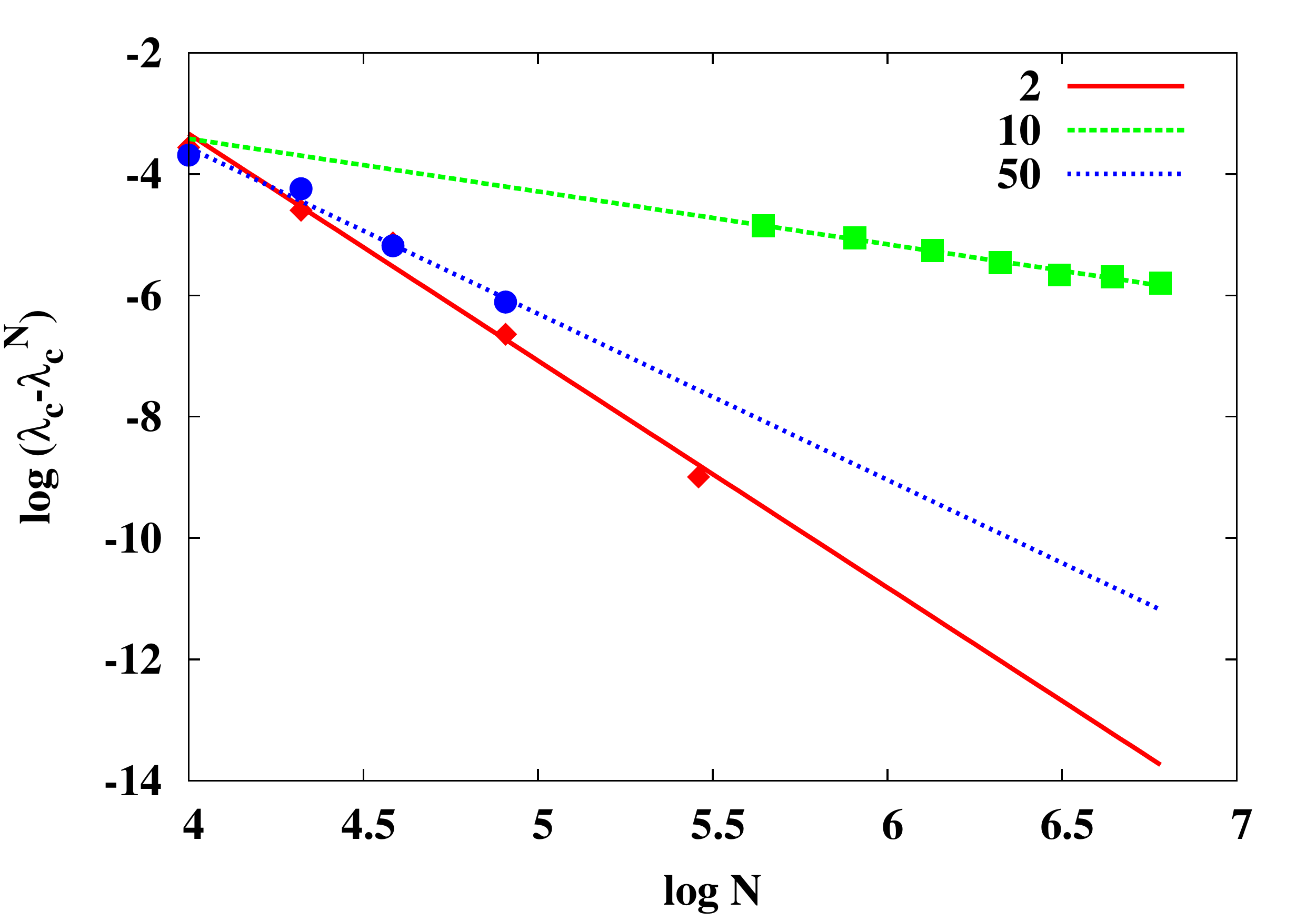}
\end{center}
\caption{(Color online.) Scaling analysis of Tsallis quantum correlation, $\cal{D}_\alpha^T$, 
for different values of $\alpha$, in the one-dimensional quantum Ising model. The legends indicate the values of $\alpha$.
Both axes are dimensionless.}
%\sout{The axes dimensions are as in Fig. \ref{fig-scaling-ren}.}
\label{fig-scaling-tsal}
\end{figure}
\begin{table}[!h]
\caption{The scaling exponents for both $\cal{D}_\alpha^R$ and $\cal{D}_\alpha^T$ for some values of $\alpha$.}
{
\begin{center}
\begin{tabular}{|c|c|c|}
\hline
$\alpha$ & $\cal{D}_\alpha^R$ & $\cal{D}_\alpha^T$\\
\hline
2.0 & -3.45 & -3.74 \\ 
10.0 & -1.28 & -0.87 \\ 
50.0 & -1.25 & -2.74  \\ 
\hline
\end{tabular}
\end{center}
}
\label{table_tt}
\end{table}
% Another importance of such  analysis is that it helps us to understand the viability of detecting the critical point in finite-sized 
% systems, which can be built by using ultracold gas systems. 
Table \ref{table_tt} exhibits the scaling exponents for both $\cal{D}_\alpha^R$ and $\cal{D}_\alpha^T$ for some values of $\alpha$. 
It is found that for $\alpha=2$, the scaling exponents are much higher for both $\cal{D}_\alpha^R$ and $\cal{D}_\alpha^T$ than
any other known measures. In particular, the 
%, e.g., when compared with the 
scaling exponents for transverse magnetization, fidelity, concurrence, quantum discord, 
and shared purity are respectively -1.69, -0.99, -1.87, -1.28, and -1.65 ~\cite{classical-Sachdev,exponent-discord,fidel_scaling,Aosterloh,shared_pur}.

\section{Conclusions}
\label{conclu}

Quantum discord is a quantum correlation measure, belonging to the information-theoretic paradigm, and it has the potential to explain
%leads to novel perspectives 
 several
quantum phenomena that 
cannot be explained by invoking the concept of quantum entanglement.
%are independent of the concept of quantum entanglement. 
In this paper, we have defined quantum correlations with generalized 
classes of entropies, viz. 
the R\'{e}nyi and the Tsallis ones. The usual quantum discord incorporates the von Neumann entropy in its definition. 
We first defined the generalized mutual information
in terms of sandwiched relative entropy distances. Using this definition of generalized mutual information, we introduced the generalized quantum correlations, 
and have shown that
they fulfill the intuitively satisfactory properties of quantum correlation measures. 
We have evaluated the generalized quantum correlations for pure states and some paradigmatic classes of mixed states.

As an application, we find that the generalized quantum correlations can detect quantum phase transitions in the transverse quantum Ising model. Interestingly,
a finite-size scaling analysis reveals that the scaling exponents obtained for the generalized quantum correlations can be significantly higher than the 
usual quantum discord as well as other order parameters, like transverse magnetization and concurrence,  at the same critical point. 
This aspect can lead to the usefulness of these measures in quantum simulators in ultracold gas experiments, potentially realizing 
finite versions of quantum spin models.
Moreover, while the derivative of the quantum discord provides only a point of inflexion at the quantum critical
point, the derivative of the generalized quantum correlations defined here signals the same critical point via a divergence.

 \begin{acknowledgments}
%AM acknowledges Mehedi Masud for useful discussions regarding numerical simulations. 
We thank  Nilanjana Datta, Mehedi Masud, Amit K. Pal, A.K. Rajagopal, and Aninda Sinha for discussions, and
acknowledge computations performed at the cluster computing facility at the Harish-Chandra Research Institute, India. 
% (http://cluster.hri.res.in/).
\end{acknowledgments}


\begin{thebibliography}{999}
\bibitem{hor} R. Horodecki, P. Horodecki, M. Horodecki, and K. Horodecki , Rev. Mod. Phys. {\bf81}, 865 (2009).
\bibitem{kavan} K. Modi, A. Brodutch, H. Cable, T. Paterek, and V.Vedral, Rev. Mod. Phys. {\bf84}, 1655 (2012).
\bibitem{aus} A. Sen(De) and U. Sen, Phys. News {\bf40}, 17 (2010).
\bibitem{briegel} H.J. Briegel, Nat. Phys. {\bf5}, 19 (2009).
\bibitem{maciej} M. Lewenstein, A. Sanpera, V. Ahuﬁnger, B. Damski, A. Sen(De), and U. Sen, Adv. Phys. {\bf56}, 243 (2007).
%\bibitem{hor} R. Horodecki et al.,Rev. Mod. Phys. {\bf81}, 865 (2009).
\bibitem{faziormp} L. Amico, R. Fazio, A. Osterloh, and V. Vedral, Rev. Mod. Phys. {\bf80}, 517 (2008). 

%\bibitem{eof} C. H. Bennett, D. P. DiVincenzo, J. A. Smolin, and W. K. Wootters, Phys. Rev. A \textbf{54}, 3824 (1996).

%\bibitem{distillable} E. M. Rains, Phys. Rev. A \textbf{60}, 173 (1999); E. M. Rains, Phys. Rev. A \textbf{60}, 179 (1999).
%; I. Devetak and A. Winter, Proc. R. Soc. A \textbf{461}, 207 (2005).
%\bibitem{VPRK} V. Vedral, M. B. Plenio, M. A. Rippin, and P. L. Knight, Phys. Rev. Lett. {\bf 78}, 2275 (1997); V. Vedral and M. B. Plenio, Phys. Rev. A {\bf 57}, 1619 (1998).
%\bibitem{VidalWerner}  G. Vidal and R. F. Werner, Phys. Rev. A \textbf{65}, 032314 (2002).
%\bibitem{hor} R. Horodecki, P. Horodecki, M. Horodecki, and K. Horodecki, Rev. Mod. Phys. \textbf{81}, 865 (2009).


%Quantifying entanglement. 

\bibitem{Chbennett3} C.H. Bennett, D.P. DiVincenzo, C.A. Fuchs, T. Mor, E. Rains, P.W. Shor, J.A. Smolin, and W.K. Wootters,
Phys. Rev. A {\bf 59}, 1070 (1999); 
C.H. Bennett, D.P. DiVincenzo, T. Mor, P.W. Shor, J.A. Smolin, and B.M. Terhal, Phys. Rev. Lett. {\bf 82}, 5385 (1999);
D.P. DiVincenzo, T. Mor, P.W. Shor, J.A. Smolin, and B.M. Terhal,
 Commun. Math. Phys. {\bf 238}, 379 (2003).
 
\bibitem{Aperes} A. Peres and W.K. Wootters, Phys. Rev. Lett. {\bf 66}, 1119 (1991);
J. Walgate, A.J. Short, L. Hardy, and V. Vedral, \emph{ibid}. {\bf 85}, 4972 (2000);
S. Virmani, M.F. Sacchi, M.B. Plenio, and D. Markham, Phys. Lett. A {\bf 288}, 62 (2001);
Y.-X. Chen and D. Yang, Phys. Rev. A {\bf 64}, 064303 (2001); \emph{ibid.} {\bf 65}, 022320 (2002); 
J. Walgate and L. Hardy, Phys. Rev. Lett. {\bf 89}, 147901 (2002);
M. Horodecki, A. Sen(De), U. Sen, and K. Horodecki, \emph{ibid}. {\bf 90}, 047902 (2003);
W.K. Wootters, Int. J. Quantum Inf. {\bf 4}, 219 (2006).
 
\bibitem{laflam} E. Knill and R. Laflamme, Phys. Rev. Lett. {\bf81}, 5672 (1998).
\bibitem{adatta}  A. Datta, A. Shaji, and C.M. Caves, Phys. Rev. Lett. {\bf100}, 050502 (2008).
\bibitem{qc}  S.L. Braunstein, C.M. Caves, R. Jozsa, N. Linden, S. Popescu, and R. Schack, Phys. Rev. Lett. {\bf83}, 1054 (1999); 
D.A. Meyer, \emph{ibid}. {\bf85}, 2014 (2000); S.L. Braunstein and A.K. Pati,
Quantum Inf. Comput. 2, {\bf399} (2002); A. Datta, S.T. Flammia, and C.M. Caves, Phys. Rev. A {\bf72}, 042316 (2005); A. Datta and
G. Vidal, \emph{ibid}. {\bf75}, 042310 (2007); B.P. Lanyon, M. Barbieri,
M.P. Almeida, and A.G. White, Phys. Rev. Lett. {\bf101}, 200501,(2008).

\bibitem{hv}L. Henderson and V. Vedral, J. Phys. A {\bf 34}, 6899 (2001).
\bibitem{oz} H. Ollivier and W.H. Zurek, Phys. Rev. Lett. {\bf 88}, 017901 (2002).


\bibitem{workdeficit} J. Oppenheim, M. Horodecki, P. Horodecki, and R. Horodecki, Phys. Rev. Lett. \textbf{89}, 180402 (2002); 
M. Horodecki, K. Horodecki, P. Horodecki, R. Horodecki, J. Oppenheim, A. Sen(De), and U. Sen, \emph{ibid.} \textbf{90}, 100402 (2003); 
I. Devetak, Phys. Rev. A {\bf 71}, 062303 (2005);
M. Horodecki, P. Horodecki, R. Horodecki, J. Oppenheim, A. Sen(De), U. Sen, and B. Synak-Radtke, Phys. Rev. A \textbf{71}, 062307 (2005).


\bibitem{minonlcl} S. Luo and S. Fu, Phys. Rev. Lett. {\bf106}, 120401 (2011).
\bibitem{qdefi}A.K. Rajagopal and R.W. Rendell, Phys. Rev. A {\bf66}, 022104 (2002); A.R. Usha Devi
and A.K. Rajagopal, Phys. Rev. Lett. {\bf100}, 140502 (2008).
\bibitem{infprara} K. Modi, T. Paterek, W. Son, V. Vedral, and M. Williamson, Phys. Rev. Lett. {\bf104}, 080501 (2010); 
I. Chakrabarty, P. Agrawal, and  A.K. Pati , Eur. Phys. J. D.  {\bf65}, 605 (2011);
 W.H. Zurek, Phys. Rev. A {\bf67}, 012320 (2003);
  M. Piani, P. Horodecki, and R. Horodecki,  Phys. Rev.Lett. {\bf100}, 090502 (2008);
S. Wu, U. V. Poulsen, and K. M{\o}lmer, Phys. Rev. A {\bf80}, 032319 (2009);
 D. Girolami, M. Paternostro, and G. Adesso, J. Phys.A: Math. Theor. {\bf44}, 352002 (2011);
 G. Adesso, and A. Datta, Phys. Rev. Lett. {\bf105}, 030501 (2010);
 P. Giorda, and M. G. A. Paris, Phys. Rev. Lett. {\bf105}, 020503 (2010);
B. Daki\'{c}, V. Vedral, and C. Brukner, Phys. Rev. Lett. {\bf105}, 190502 (2010);
 S. Luo, S. Fu, and N. Li, Phys. Rev. A {\bf82}, 052122 (2010);
 B. Aaronson, R.L. Franco, G. Compagno, and G. Adesso,  New J. Phys. {\bf15}, 093022 (2013);
 B. Bellomo, G.L. Giorgi, F. Galve, R. Lo Franco, G. Compagno, and R. Zambrini, Phys. Rev. A {\bf85}, 032104 (2012).

 \bibitem{wherl} A. Wehrl, Rev. Mod. Phys. {\bf50}, 221 (1978).
\bibitem{von} J. von Neumann, \emph{Mathematical Foundations of Quantum Me-
chanics} (Princeton University Press, Princeton, 1955)(Eng Translation by R.T. Beyer).

\bibitem{renyi} A. R\'{e}nyi. \emph{On measures of information and entropy}. In Proc. Symp. on Math., Stat. and Probability, 547 
(University of California Press, Berkeley, 1961).
\bibitem{tsallis} C. Tsallis, J. Stat. Phys. {\bf52}, 479 (1988); C. Tsallis, R. S. Mendes, and A. R. Plastino, Physica A {\bf261}, 534 (1998).
\bibitem{chernf} K.M.R. Audenaert, J. Calsamiglia, R. Munoz-Tapia, E. Bagan, Ll. Masanes, A. Acin, and F. Verstraete. Phys. Rev. Lett. {\bf98}, 160501 (2007).
\bibitem{mosonyi}M. Mosonyi and F. Hiai. IEEE Trans. Inf. Th. {\bf57}, 2474 (2011).
\bibitem{holography} M. Headrick, Phys. Rev.D {\bf82}, 126010 (2010);
H. Casini, M. Huerta, and R.C. Myers, JHEP {\bf05}, 036 (2011);
L.-Y. Hung, R.C. Myers, M. Smolkin, and Alexandre Yale, \emph{ibid}. {\bf12}, 047 (2011).
\bibitem{conmat}F. Franchini, A.R. Its, and V.E. Korepin, J. Phys. A: Math. Theor. {\bf41}  025302 (2008);
A.R. Its and V.E. Korepin, Theor. and Math. Phys. (Springer) {\bf164} 1136 (2010);
H. Li and F.D.M. Haldane, Phys. Rev. Lett. {\bf101}, 010504 (2008);
S.T. Flammia, A. Hamma, T.L. Hughes, and X.-G. Wen, Phys. Rev. Lett. {\bf103}, 261601 (2009);
M.A. Metlitski, C.A. Fuertes, and S. Sachdev, Phys. Rev. B {\bf80}, 115122 (2009);
 M.B. Hastings, I. Gonz\'{a}lez, A.B. Kallin, and R.G. Melko, Phys. Rev. Lett. {\bf104}, 157201 (2010).

\bibitem{abeakrqe}S. Abe and A.K. Rajagopal, Phys. Rev. A {\bf60}, 3461 (1999).
\bibitem{abeakrlocal}S. Abe and A.K. Rajagopal, Physica A {\bf289}, 157 (2001).
\bibitem{gunheml}O. G\"{u}hne and M. Lewenstein, Phys. Rev. A {\bf70}, 022316 (2004).
\bibitem{tsallisref}S. Abe and A.K. Rajagopal, Phys. Rev. Lett. {\bf83}, 1711 (1999); T. Yamano, Phys. Rev. E {\bf63}, 046105 (2001);  
J.S. Kim, Phys. Rev. A {\bf81}, 062328 (2010).
\bibitem{statmech} C. Tsallis, \emph{Introduction to Nonextensive Statistical Mechanics}, (Springer, Ney York, 2009);
C. Beck and F. Sch\..{o}gl, \emph{Thermodynamics of Chaotic Systems: An Introduction}, (Cambridge University Press, Cambridge, 1993);
P. Jizba and T. Arimitsu, Annals Phys. {\bf312} 17 (2004).
\bibitem{dakic} B. Daki\'{c}, V. Vedral, and C. Brukner, Phys. Rev. Lett. {\bf105}, 190502 (2010).
\bibitem{tradtsallis} S.Abe, Phys.Rev.A {\bf68}, 032302 (2003); S.Abe, Phys. Lett. A {\bf312}, 336 (2003).
\bibitem{wilde} M.M. Wilde, A. Winter, and D. Yang, arXiv:1306.1586 (2013).
\bibitem{lennert} M. M\"{u}ller-Lennert, F. Dupuis, O. Szehr, S. Fehr, and M. Tomamichel, arXiv:1306.3142 (2013).

\bibitem{ogawa} M. Mosonyi and T. Ogawa. arXiv:1309.3228 (2013).
\bibitem{akr}A.K. Rajagopal, Sudha, A.S Nayak, and A.R. Usha Devi, Phys. Rev. A {\bf89}, 012331 (2014).
\bibitem{rennerphd} R. Renner, \emph{Security of quantum key distribution}, PhD thesis. arXiv:quant-ph/0512258 (2005).
\bibitem{nilans}N. Datta, IEEE Trans. Inf. Th. {\bf55}, 2816–2826 (2009).
\bibitem{renner1} F. Dupuis, L. Kraemer, P. Faist, J.M. Renes, and R. Renner, XVIIth International Congress on Mathematical Physics, 134 (2013).

\bibitem{buscemi}  F. Buscemi and N. Datta, Phys. Rev. Lett. {\bf106}, 130503 (2011).
\bibitem{wang}L. Wang and R. Renner, Phys. Rev. Lett. {\bf108}, 200501 (2012).
\bibitem{toma} M. Tomamichel, \emph{A Framework for Non-Asymptotic Quantum Information Theory}. PhD Thesis. arXiv:1203.2142 (2012).
\bibitem{frustration}U. Marzolino, S.M. Giampaolo, and F. Illuminati, Phys. Rev. A {\bf88}, 020301(R) (2013).
\bibitem{renyiq}R. Horodecki, P. Horodecki, and M. Horodecki, Phys.Lett. A. {\bf210}, 377 (1996); R. Horodecki and M. Horodecki, Phys. Rev. A {\bf54}, 1838 (1996).
\bibitem{tsallisq} S. Abe and A.K. Rajagopal, Physica A {\bf289}, 157 (2001).
\bibitem{gen_q_entropy}X. Hu  and Z. Ye, J. Math. Phys. {\bf47}, 023502 (2006), and references therein.
\bibitem{umegaki} H.Umegaki, Kodai Math. Sem. Rep. {\bf14} 59 (1962); G. Lindblad, Commun Math. Phys. {\bf33}, 305 (1973).





\bibitem{lieb} R.L. Frank and E.H. Lieb, arXiv:1306.5358 (2013).

\bibitem{beigi} S. Beigi, J. Math. Phys. {\bf 54}, 122202 (2013).
\bibitem{rastegin}A.E. Rastegin, arXiv:1309:6048 (2013).

\bibitem{nilan} K.M.R. Audenaert and N. Datta,  arXiv:1310.7178 (2013).

%XVIIth Int’l Conf. on Mathematical Physics, Aalborg (2012). See also arXiv:1211.3141 (2012).

\bibitem{qmi} W.H. Zurek, in {\it Quantum Optics, Experimental Gravitation and Measurement Theory},
edited by P. Meystre and M.O. Scully (Plenum, New York, 1983); S.M. Barnett and S.J.D. Phoenix, Phys. Rev. A {\bf 40}, 2404 (1989).

\bibitem{Cerf} N.J. Cerf and C. Adami, Phys. Rev. Lett. \textbf{79}, 5194 (1997).

\bibitem{GROIS} B. Schumacher and M.A. Nielsen, Phys. Rev. A \textbf{54}, 2629 (1996); 
B. Groisman, S. Popescu, and A. Winter, Phys. Rev. A \textbf{72}, 032317 (2005).
\bibitem{Barry} S. Bandyopadhyay, G. Gour, and B.C. Sanders,
J. Math. Phys. \textbf{48}, 012108 (2007). 
%\bibitem{vedre} V. Vedral, M. B. Plenio, M. A. Rippin and P. L. Knight, Phys. Rev. Lett. {\bf78}, 2275–2279 (1997).
\bibitem{tsallisdis3} A.P. Majtey, A.R. Plastino, A. Plastino, Physica A {\bf391}, 2491 (2012).
\bibitem{tsallisdis1} J. Jurkowski, Int. J. Quantum Inform. {\bf11}, 1350013 (2013).
\bibitem{tsallisdis2} D.P. Chi, J.S. Kim and K. Lee, arXiv:1302.2984 (2013).
\bibitem{tsallisdis4} A.C.S. Costa and R.M. Angelo, Phys. Rev. A {\bf87}, 032109 (2013).
\bibitem{renyidis} G. Adesso, D. Girolami, and A. Serafini, Phys. Rev. Lett. {\bf109}, 190502 (2012).
\bibitem{wildemutual} M. Berta, K.P. Seshadreesan, M. M. Wilde, arXiv:1403.6102v1 (2014).
\bibitem{ourpre}A. Misra, A. Biswas, A. K. Pati, A. Sen(De), and U. Sen, Phys. Rev. E \textbf{91}, 052125 (2015).
%\bibitem{wildepub} M. Berta, K.P. Seshadreesan, M. M. Wilde, J. Math. Phys. {\bf56}, 022205 (2015).
\bibitem{wildediscord} K.P. Seshadreesan, M. Berta, M.M. Wilde, arXiv:1410.1443 (2014).
\bibitem{bound}P. Horodecki, Phys. Lett. A {\bf232}, 333 (1997); P. Horodecki, M. Horodecki, and R. Horodecki, Phys. Rev. Lett. \textbf{82}, 1056 (1999).
\bibitem{dist_entngl}C. H. Bennett, G. Brassard, S. Popescu, B. Schumacher, J. Smolin, and W. K. Wootters, Phys. Rev. Lett. {\bf76}, 722 (1996);
 C. H. Bennett, D. P. DiVincenzo, J. Smolin, and W. K. Wootters, Phys. Rev. A {\bf54}, 3824 (1996); E.M. Rains, Phys. Rev. A \textbf{60}, 173 (1999).
\bibitem{furuchi} S. Furuichi, K. Yanagi, K. Kuriyama, J.Math.Phys. {\bf45}, 4868 (2004).
\bibitem{traddpi}E.H. Lieb, Adv. Math. {\bf11}, 267 (1973); A. Uhlmann, Commun. Math. Phys. {\bf54}, 21 (1977); D. Petz, Rep. Math. Phys. {\bf23}, 57 (1986);
M. Hayashi,\emph{ Quantum Information Theory: An Introduction} (Springer, 2006).
\bibitem{Barouch-McCoy} E. Lieb, T. Schultz, and D. Mattis, Ann. Phys. \textbf{16}, 407 (1961);
E. Barouch, B.M. McCoy, and M. Dresden, Phys. Rev. \textbf{2}, 1075 (1970);
E. Barouch and B.M. McCoy, Phys. Rev. \textbf{3}, 786 (1971);
S. Sachdev, {\em Quantum Phase Transistions} (Cambridge University Press, Cambridge, 2011).


\bibitem{Strotzky} D. Porras and J.I. Cirac, Laser Physics {\bf 15}, 88 (2004);
A. Friedenauer, H. Schmitz, J.T. Gl{\" u}ckert, D. Porras, and T. Sch\"{a}tz, arXiv:0802.4072 (2008);
H. H\"{a}ffner, C.F. Roosa, and R. Blatt, Physics Reports {\bf 469}, 155 (2008);
 K. Kim, M.-S. Chang, S. Korenblit, R. Islam, E.E. Edwards, J.K. Freericks, G.-D. Lin, L.-M. Duan, and C. Monroe, Nature {\bf 465}, 590 (2010);
 S. Trotzky, Y.-A. Chen, U. Schnorrberger, P. Cheinet, and I. Bloch,
Phys. Rev. Lett. \textbf{105}, 265303 (2010); M. Cramer, M. B. Plenio, and H. Wunderlich,
\emph{ibid.} \textbf{106}, 020401 (2011); P.J. Pemberton-Ross and A. Kay,
\emph{ibid.}, 020503 (2011);
P. Richerme, C. Senko, J. Smith, A. Lee, S. Korenblit, and C. Monroe, arXiv:1305.2253 (2013), and references therein.




% \bibitem{Strotzky}  S. Trotzky, Y.-A. Chen, U. Schnorrberger, P. Cheinet, and I. Bloch,
% Phys. Rev. Lett. \textbf{105}, 265303 (2010); M. Cramer, M. B. Plenio, and H. Wunderlich,
% \emph{ibid.} \textbf{106}, 020401 (2011); P.J. Pemberton-Ross and A. Kay,
% \emph{ibid.}, 020503 (2011), and references therein.


\bibitem{Mschechter} D. Bitko, T.F. Rosenbaum, and  G. Aeppli, Phys. Rev. Lett. {\bf 77}, 940 (1996);
Ch. R\"{u}egg, B. Normand, M. Matsumoto, A. Furrer, D.F. McMorrow, K.W. Krämer, H.-U. G\"{u}del, S.N. Gvasaliya, H. Mutka, and M. Boehm, 
\emph{ibid.} {\bf 100}, 205701 (2008); 
M. Schechter and P.C.E. Stamp, Phys. Rev. B {\bf 78}, 054438 (2008);
R. Coldea, D.A. Tennant, E.M. Wheeler, E. Wawrzynska, D. Prabhakaran, M. Telling, K. Habicht, P. Smeibidl, and K. Kiefer, Science {\bf 327}, 177 (2010), and references therein.


% \bibitem{Mschechter} M. Schechter and P. C. E. Stamp, Phys. Rev. B {\bf 78}, 054438 (2008);
% R. Coldea, D. A. Tennant, E. M. Wheeler, E. Wawrzynska, D. Prabhakaran, M. Telling, K. Habicht, P. Smeibidl, and K. Kiefer, Science {\bf 327}, 177 (2010).

\bibitem{Mefisher} M.E. Fisher, Physica {\bf 26}, 618 (1960); S. Katsura, Phys. Rev. {\bf 127}, 1508 (1962). 
\bibitem{Aosterloh} A. Osterloh, L. Amico, G. Falci, and R. Fazio, Nature \textbf{416}, 618 (2002); 
T.J. Osborne and M.A. Nielsen, Phys. Rev. A \textbf{66}, 032110 (2002).
\bibitem{Ashimony} A. Shimony, Ann. N.Y. Acad. Sci. \textbf{755}, 675 (1995); 
H. Barnum and N. Linden, J. Phys. A \textbf{34}, 6787 (2001);
T.-C. Wei and P.M. Goldbart, Phys. Rev. A \textbf{68}, 042307 (2003);
 T.-C. Wei, D. Das, S. Mukhopadyay, S. Vishveshwara, and P.M. Goldbart , Phys. Rev. A {\bf 71}, 060305(R) (2005);
R. Or\'{u}s, Phys. Rev. Lett. {\bf 100}, 130502 (2008);
R. Or\'{u}s, S. Dusuel, and J. Vidal, Phys. Rev. Lett. {\bf 101}, 025701 (2008);
 M. Balsone, F. Dell’Anno, S. De Siena, and F. Illuminati, Phys. Rev. A \textbf{77}, 062304 (2008);
R. Or\'{u}s, \emph{ibid.} {\bf 78}, 062332 (2008);
Q.-Q. Shi, R. Or\'{u}s, J.O. Fj\ae{}restad, and H.-Q. Zhou, New J. Phys. {\bf 12}, 025008 (2010);
R. Or\'{u}s and T.-C. Wei, Phys. Rev. B {\bf 82}, 155120 (2012).

\bibitem{Asde} A. Sen(De) and U. Sen,  Phys. Rev. A {\bf 81}, 012308 (2010).
\bibitem{Abiswas} A. Biswas, R. Prabhu, A. Sen(De), and U. Sen, Phys. Rev. A \textbf{90}, 032301 (2014).
\bibitem{Rdillenschneider} R. Dillenschneider, Phys. Rev. B \textbf{78}, 224413 (2008);
A. Dutta, U. Divakaran, D. Sen, B.K. Chakrabarti, T.F. Rosenbaum, and G. Aeppli, arXiv:1012.0653 (2010);
M.S. Sarandy, Phys. Rev. A {\bf 80}, 022108 (2009); Y. Huang, Phys. Rev. B \textbf{89}, 054410 (2014), and references therein.


\bibitem{Lidar-qpt}L.-A. Wu, M.S. Sarandy, and D.A. Lidar,
Phys. Rev. Lett. \textbf{93}, 250404 (2004).

%\bibitem{sarandy} M.S. Sarandy, Phys. Rev. A {\bf 80}, 022108 (2009).


\bibitem{Rislam} R. Islam, E.E. Edwards, K. Kim, S. Korenblit, C. Noh, H. Carmichael, G.-D.Lin, L.-M. Duan, 
C.-C. Joseph Wang, J.K. Freericks, and C. Monroe,  Nature Commun. {\bf2}, 377 (2011) and references therein.




\bibitem{exponent-discord} B. Tomasello, D. Rossini, A. Hamma, and L. Amico, Europhys. Lett. \textbf{96},  27002 (2011); 
Int. Jour. Mod. Phys. B \textbf{26},  1243002 (2012).

\bibitem{classical-Sachdev} J.J. Binney, N.J. Dowrick,  A.J. Fisher, and  M.E.J. Newman,   
\emph{The Theory of Critical Phenomena: An Introduction to the 
Renormalization Group} (Clarendon, Oxford, 1992).


\bibitem{fidel_scaling} H.-Q.Zhou, H.-H. Zhao, and B. Li, J. Phys. A:Math. Theor. \textbf{41}, 492002 (2008).



\bibitem{shared_pur} A. Biswas, A. Sen (De), and U. Sen, Phys. Rev. A \textbf{89}, 032331 (2014).





\end{thebibliography}
\end{document}